\documentclass[%
 reprint,
superscriptaddress,
 amsmath,amssymb,
 aps,
]{revtex4-2}

\usepackage[dvipdfmx]{graphicx}
\usepackage{dcolumn}
\usepackage{bm}
\usepackage{color}
\usepackage{hyperref}
\usepackage{physics}
\usepackage{orcidlink}

\newcommand{\software}[1]{{\tt #1}}
\newcommand{\mapcube}{\langle\mathcal{M}_{\rm ap}^3\rangle}
\usepackage{lineno}
\usepackage{afterpage}

\begin{document}

\preprint{APS/123-QED}

\title{Cosmology with second and third-order shear statistics for the Dark Energy Survey: Methods and simulated analysis}

\author{R. C. H. ~Gomes\orcidlink{0000-0002-3800-5662}}
\affiliation{Department of Physics and Astronomy, University of Pennsylvania, Philadelphia, PA 19104, USA}

\author{S. ~Sugiyama}
\affiliation{Department of Physics and Astronomy, University of Pennsylvania, Philadelphia, PA 19104, USA}

\author{B. ~Jain}
\affiliation{Department of Physics and Astronomy, University of Pennsylvania, Philadelphia, PA 19104, USA}

\author{M. ~Jarvis}
\affiliation{Department of Physics and Astronomy, University of Pennsylvania, Philadelphia, PA 19104, USA}

\author{D. ~Anbajagane}
\affiliation{Department of Astronomy and Astrophysics, University of Chicago, Chicago, IL 60637, USA}

\author{M. ~Gatti}
\affiliation{Kavli Institute for Cosmological Physics, University of Chicago, Chicago, IL 60637, USA}

\author{D. ~Gebauer}
\affiliation{Universitäts-Sternwarte, Fakultät für Physik, Ludwig-Maximilians Universität München,
Scheinerstraße 1, 81679 München, Germany}
\affiliation{Max Planck Institute for Extraterrestrial Physics, Giessenbachstraße 1, 85748 Garching,
Germany}

\author{Z. ~Gong}
\affiliation{Universitäts-Sternwarte, Fakultät für Physik, Ludwig-Maximilians Universität München,
Scheinerstraße 1, 81679 München, Germany}
\affiliation{Max Planck Institute for Extraterrestrial Physics, Giessenbachstraße 1, 85748 Garching,
Germany}

\author{A. ~Halder}
\affiliation{Institute of Astronomy and Kavli Institute for Cosmology, University of Cambridge, Madingley Road,
Cambridge, CB3 0HA, United Kingdom}
\affiliation{Jesus College, Jesus Lane,
Cambridge, CB5 8BL, United Kingdom}
\affiliation{Max Planck Institute for Extraterrestrial Physics, Giessenbachstraße 1, 85748 Garching,
Germany}
\affiliation{University Observatory, Ludwig-Maximilians-University of Munich, Scheinerstraße 1, 81679 Munich, Germany}

\author{G. A. ~Marques}
\affiliation{Fermi National Accelerator Laboratory, Batavia, IL 60510, USA}
\affiliation{Kavli Institute for Cosmological Physics, University of Chicago, Chicago, IL 60637, USA}

\author{S. ~Pandey}
\affiliation{Columbia Astrophysics Laboratory, Columbia University, 550 West 120th Street, New York, NY 10027, USA}

\author{J.~L.~Marshall}
\affiliation{George P. and Cynthia Woods Mitchell Institute for Fundamental Physics and Astronomy, and Department of Physics and Astronomy, Texas A\&M University, College Station, TX 77843, USA}

\author{S.~Allam}
\affiliation{Fermi National Accelerator Laboratory, P. O. Box 500, Batavia, IL 60510, USA}

\author{O.~Alves}
\affiliation{Department of Physics, University of Michigan, Ann Arbor, MI 48109, USA}

\author{F.~Andrade-Oliveira}
\affiliation{Physik-Institut, University of Zürich, Winterthurerstrasse 190, CH-8057 Zürich, Switzerland}

\author{D.~Bacon}
\affiliation{Institute of Cosmology and Gravitation, University of Portsmouth, Portsmouth, PO1 3FX, UK}

\author{J.~Blazek}
\affiliation{Department of Physics, Northeastern University, Boston, MA 02115, USA}

\author{S.~Bocquet}
\affiliation{University Observatory, Faculty of Physics, Ludwig-Maximilians-Universit\"at, Scheinerstr. 1, 81679 Munich, Germany}

\author{D.~Brooks}
\affiliation{Department of Physics \& Astronomy, University College London, Gower Street, London, WC1E 6BT, UK}

\author{A.~Carnero~Rosell}
\affiliation{Instituto de Astrofisica de Canarias, E-38205 La Laguna, Tenerife, Spain}
\affiliation{Laborat\'orio Interinstitucional de e-Astronomia - LIneA, Av. Pastor Martin Luther King Jr, 126 Del Castilho, Nova Am\'erica Offices, Torre 3000/sala 817 CEP: 20765-000, Brazil}
\affiliation{Universidad de La Laguna, Dpto. Astrof\'{\i}sica, E-38206 La Laguna, Tenerife, Spain}

\author{J.~Carretero}
\affiliation{Institut de F\'{\i}sica d'Altes Energies (IFAE), The Barcelona Institute of Science and Technology, Campus UAB, 08193 Bellaterra (Barcelona) Spain}

\author{L.~N.~da Costa}
\affiliation{Laborat\'orio Interinstitucional de e-Astronomia - LIneA, Av. Pastor Martin Luther King Jr, 126 Del Castilho, Nova Am\'erica Offices, Torre 3000/sala 817 CEP: 20765-000, Brazil}

\author{P.~Doel}
\affiliation{Department of Physics \& Astronomy, University College London, Gower Street, London, WC1E 6BT, UK}

\author{C.~Doux}
\affiliation{Department of Physics and Astronomy, University of Pennsylvania, Philadelphia, PA 19104, USA}
\affiliation{Universit\'e Grenoble Alpes, CNRS, LPSC-IN2P3, 38000 Grenoble, France}

\author{S.~Everett}
\affiliation{California Institute of Technology, 1200 East California Blvd, MC 249-17, Pasadena, CA 91125, USA}

\author{B.~Flaugher}
\affiliation{Fermi National Accelerator Laboratory, P. O. Box 500, Batavia, IL 60510, USA}

\author{J.~Frieman}
\affiliation{Department of Astronomy and Astrophysics, University of Chicago, Chicago, IL 60637, USA}
\affiliation{Fermi National Accelerator Laboratory, P. O. Box 500, Batavia, IL 60510, USA}
\affiliation{Kavli Institute for Cosmological Physics, University of Chicago, Chicago, IL 60637, USA}

\author{J.~Garc\'ia-Bellido}
\affiliation{Instituto de Fisica Teorica UAM/CSIC, Universidad Autonoma de Madrid, 28049 Madrid, Spain}

\author{E.~Gaztanaga}
\affiliation{Institut d'Estudis Espacials de Catalunya (IEEC), 08034 Barcelona, Spain}
\affiliation{Institute of Cosmology and Gravitation, University of Portsmouth, Portsmouth, PO1 3FX, UK}
\affiliation{Institute of Space Sciences (ICE, CSIC),  Campus UAB, Carrer de Can Magrans, s/n,  08193 Barcelona, Spain}

\author{D.~Gruen}
\affiliation{University Observatory, Faculty of Physics, Ludwig-Maximilians-Universit\"at, Scheinerstr. 1, 81679 Munich, Germany}

\author{R.~A.~Gruendl}
\affiliation{Center for Astrophysical Surveys, National Center for Supercomputing Applications, 1205 West Clark St., Urbana, IL 61801, USA}
\affiliation{Department of Astronomy, University of Illinois at Urbana-Champaign, 1002 W. Green Street, Urbana, IL 61801, USA}

\author{G.~Gutierrez}
\affiliation{Fermi National Accelerator Laboratory, P. O. Box 500, Batavia, IL 60510, USA}

\author{K.~Herner}
\affiliation{Fermi National Accelerator Laboratory, P. O. Box 500, Batavia, IL 60510, USA}

\author{S.~R.~Hinton}
\affiliation{School of Mathematics and Physics, University of Queensland,  Brisbane, QLD 4072, Australia}

\author{D.~L.~Hollowood}
\affiliation{Santa Cruz Institute for Particle Physics, Santa Cruz, CA 95064, USA}

\author{K.~Honscheid}
\affiliation{Center for Cosmology and Astro-Particle Physics, The Ohio State University, Columbus, OH 43210, USA}
\affiliation{Department of Physics, The Ohio State University, Columbus, OH 43210, USA}

\author{D.~Huterer}
\affiliation{Department of Physics, University of Michigan, Ann Arbor, MI 48109, USA}

\author{D.~J.~James}
\affiliation{Center for Astrophysics $\vert$ Harvard \& Smithsonian, 60 Garden Street, Cambridge, MA 02138, USA}

\author{N.~Jeffrey}
\affiliation{Department of Physics \& Astronomy, University College London, Gower Street, London, WC1E 6BT, UK}

\author{J.~Mena-Fern{\'a}ndez}
\affiliation{LPSC Grenoble - 53, Avenue des Martyrs 38026 Grenoble, France}

\author{R.~Miquel}
\affiliation{Instituci\'o Catalana de Recerca i Estudis Avan\c{c}ats, E-08010 Barcelona, Spain}
\affiliation{Institut de F\'{\i}sica d'Altes Energies (IFAE), The Barcelona Institute of Science and Technology, Campus UAB, 08193 Bellaterra (Barcelona) Spain}

\author{J.~Muir}
\affiliation{Department of Physics, University of Cincinnati, Cincinnati, Ohio 45221, USA}
\affiliation{Perimeter Institute for Theoretical Physics, 31 Caroline St. North, Waterloo, ON N2L 2Y5, Canada}

\author{R.~L.~C.~Ogando}
\affiliation{Observat\'orio Nacional, Rua Gal. Jos\'e Cristino 77, Rio de Janeiro, RJ - 20921-400, Brazil}

\author{M.~E.~S.~Pereira}
\affiliation{Hamburger Sternwarte, Universit\"{a}t Hamburg, Gojenbergsweg 112, 21029 Hamburg, Germany}

\author{A.~Pieres}
\affiliation{Laborat\'orio Interinstitucional de e-Astronomia - LIneA, Av. Pastor Martin Luther King Jr, 126 Del Castilho, Nova Am\'erica Offices, Torre 3000/sala 817 CEP: 20765-000, Brazil}
\affiliation{Observat\'orio Nacional, Rua Gal. Jos\'e Cristino 77, Rio de Janeiro, RJ - 20921-400, Brazil}

\author{A.~A.~Plazas~Malag\'on}
\affiliation{Kavli Institute for Particle Astrophysics \& Cosmology, P. O. Box 2450, Stanford University, Stanford, CA 94305, USA}
\affiliation{SLAC National Accelerator Laboratory, Menlo Park, CA 94025, USA}

\author{S.~Samuroff}
\affiliation{Department of Physics, Northeastern University, Boston, MA 02115, USA}
\affiliation{Institut de F\'{\i}sica d'Altes Energies (IFAE), The Barcelona Institute of Science and Technology, Campus UAB, 08193 Bellaterra (Barcelona) Spain}

\author{E.~Sanchez}
\affiliation{Centro de Investigaciones Energ\'eticas, Medioambientales y Tecnol\'ogicas (CIEMAT), Madrid, Spain}

\author{D.~Sanchez Cid}
\affiliation{Centro de Investigaciones Energ\'eticas, Medioambientales y Tecnol\'ogicas (CIEMAT), Madrid, Spain}
\affiliation{Physik-Institut, University of Zürich, Winterthurerstrasse 190, CH-8057 Zürich, Switzerland}

\author{B.~Santiago}
\affiliation{Instituto de F\'\i sica, UFRGS, Caixa Postal 15051, Porto Alegre, RS - 91501-970, Brazil}
\affiliation{Laborat\'orio Interinstitucional de e-Astronomia - LIneA, Av. Pastor Martin Luther King Jr, 126 Del Castilho, Nova Am\'erica Offices, Torre 3000/sala 817 CEP: 20765-000, Brazil}

\author{I.~Sevilla-Noarbe}
\affiliation{Centro de Investigaciones Energ\'eticas, Medioambientales y Tecnol\'ogicas (CIEMAT), Madrid, Spain}

\author{M.~Smith}
\affiliation{Physics Department, Lancaster University, Lancaster, LA1 4YB, UK}

\author{E.~Suchyta}
\affiliation{Computer Science and Mathematics Division, Oak Ridge National Laboratory, Oak Ridge, TN 37831}

\author{M.~E.~C.~Swanson}
\affiliation{Center for Astrophysical Surveys, National Center for Supercomputing Applications, 1205 West Clark St., Urbana, IL 61801, USA}

\author{G.~Tarle}
\affiliation{Department of Physics, University of Michigan, Ann Arbor, MI 48109, USA}

\author{C.~To}
\affiliation{Department of Astronomy and Astrophysics, University of Chicago, Chicago, IL 60637, USA}

\author{V.~Vikram}
\affiliation{Department of Physics, Central University of Kerala, 93VR+RWF, CUK Rd, Kerala 671316, Índia}

\author{N.~Weaverdyck}
\affiliation{Department of Astronomy, University of California, Berkeley,  501 Campbell Hall, Berkeley, CA 94720, USA}
\affiliation{Lawrence Berkeley National Laboratory, 1 Cyclotron Road, Berkeley, CA 94720, USA}

\author{J.~Weller}
\affiliation{Max Planck Institute for Extraterrestrial Physics, Giessenbachstrasse, 85748 Garching, Germany}
\affiliation{Universit\"ats-Sternwarte, Fakult\"at f\"ur Physik, Ludwig-Maximilians Universit\"at M\"unchen, Scheinerstr. 1, 81679 M\"unchen, Germany}

\collaboration{DES Collaboration}

\date{\today}

\begin{abstract}
We present a new pipeline designed for the robust inference of cosmological parameters using both second- and third-order shear statistics. We build a theoretical model for rapid evaluation of three-point correlations using our \texttt{fastnc} code and integrate it into the \texttt{CosmoSIS} framework. We measure the two-point functions $\xi_{\pm}$ and the full configuration-dependent three-point shear correlation functions across all auto- and cross-redshift bins. We compress the three-point functions into the mass aperture statistic $\mapcube$ for a set of 796 simulated shear maps designed to model the Dark Energy Survey (DES) Year 3 data.  
We  estimate from it the full covariance matrix and model the effects of intrinsic alignments, shear calibration biases and photometric redshift uncertainties. We apply scale cuts to minimize the contamination from the baryonic signal as modeled through hydrodynamical simulations.
We find a significant improvement of $83\%$ on the Figure of Merit in the $\Omega_{\rm m}$-$S_8$ plane when we add the $\mapcube$ data to  $\xi_{\pm}$. We present our findings for all relevant cosmological and systematic uncertainty parameters and discuss the complementarity of third-order and second-order statistics.

\end{abstract}

\maketitle

\section{Introduction}\label{sec:introduction}

The current state of observational cosmology is one of tensions and expectations \cite{Perivolaropoulos_2022} \cite{Efstathiou.challenges}. While discrepancies continue to be identified between standard cosmological predictions and observational data, the upcoming Stage IV surveys offer a promising hope for identifying where we need to refine or go beyond the $\Lambda$CDM model and where we need to improve our understanding of observational systematics \cite{Eifler.Simet.Krause}. This context has been driving cosmologists towards the development of techniques to extract the most information out of the available data. 

For weak lensing data, going beyond the two-point correlation function of the cosmic shear field allows us to probe the field's non-Gaussian features, which can both lead us to a tightening of the parameter space constraints and offer improved characterization of systematics such as the intrinsic alignment of galaxies \cite{Pyne.Joachimi.2021}. Cosmic shear two-point analyses have been performed with DES-Y3 data \cite{Secco_2022}\cite{Amon.Weller.2021}, as well as with data from KiDS-1000 \cite{Asgari.Valentijn.2020} and HSC Y3 \cite{Dalal.Wang.2023}\cite{Li.Wang.2023}.

Many studies have been conducted using higher-order statistics of the cosmic shear field as well. 
An approach that uses moments of the weak lensing mass maps was conducted by \citet{Gatti.collaboration.2020} with Dark Energy Survey (DES) data. The use of third moments of the field (encoding non-Gaussian information) in addition to the second moments is shown to give an improvement of $15\%$ in $S_8$ constraints and of $25\%$ in $\Omega_{\rm m}$ constraints relative to the use of second moments alone. A similar level of improvement was found by \citet{Gong_2023} on simulated data when adding the integrated shear three-point function to the shear two-point correlation functions $\xi_{\pm}$. 

Additional state-of-the-art higher-order approaches are discussed and reviewed by \citet{Ajani.thesis}, and include Minkowski functionals \cite{Petri_2013}, weak lensing peak counts and minimum counts \cite{Marques_2024}, scattering transforms \cite{Cheng.Yuan-Sen}, wavelet phase harmonics \cite{Allys_2020}, homology statistics \cite{Heydenreich_2021_homology}, and PDF/CDFs \cite{anbajagane20233rdmomentpracticalstudy} \cite{barthelemy2024makingleapimodelling}. Finally, deep learning approaches at the field level have also been developed and applied to lensing mass maps \cite{Fluri.Hofmann.2018}\cite{Jeffrey.Lanusse.2020}\cite{Ribli.Pataki.2019}.

Given that the full shear three-point correlation function has been measured in DES-Y3 data with high signal-to-noise \cite{Secco.Weller.2022}, the question arises: How much additional information would this statistic provide to the two-point shear constraints from DES-Y3? Full shear three-point function measurements typically yield us data vectors with large dimensions, which are not optimal for covariance determination. However, a PCA analysis by \citet{Heydenreich.Schneider.2022} shows that the information content of the full three-point function is similar to that of the mass aperture statistic ($\mapcube$). Thus, the latter can be interpreted as a physically motivated efficient compression of the former.

The usage of $\mapcube$ has been shown by \citet{Burger.Martinet.2023} to lead to complementary constraints to those from two-point statistics. Their analysis with KiDS-1000 data provided a factor of two improvement in the joint $\Omega_{\rm m}$-$S_8$ constraints. Nonetheless, this was partially driven by the fact that their setup for the two-point analysis was less constraining than that of the fiducial KiDS-1000 cosmic shear analysis \cite{Asgari.Valentijn.2020}. In light of this result, we can expect joint $\xi_{\pm}$ and $\mapcube$ constraints to yield significant improvement relative to two-point analyses in the context of other collaborations. 

This paper is motivated by these recent developments and establishes a pipeline for the analysis of the third order cosmic shear information on DES-Y3 data. We construct a theoretical model for the mass aperture statistic, estimate its covariance through simulations, and build a robust likelihood for cosmological analyses. These developments are complemented by a companion paper, that will describe the analysis with DES-Y3 data.

The paper is organized as follows. 
In section~\ref{sec:model}, we describe our theoretical model for the three-point correlation function and the mass aperture statistic, including the modeling of observational systematics and a description of our neural network emulator. In section~\ref{sec:covariance}, we present our data vector and covariance. In section~\ref{sec:methods}, we describe the parameter inference scheme and our analysis choices. Finally, in section~\ref{sec:result}, we show the parameter estimation results of our simulated analysis, validating therefore our pipeline for use with DES data. Our concluding remarks are made in section~\ref{sec:conclusion}.

\section{Theoretical Modeling}\label{sec:model}

\subsection{Basics of Weak Lensing}\label{sec:}

The study of weak lensing allows us to probe the matter density field of the universe in an unbiased way, because it is sensitive to the total matter density including both visible and dark matter. The convergence field on the sky coordinate $\bm{X}$, which characterizes the isotropic magnification of the background galaxy's shape, is obtained by the line-of-sight integration of the matter density contrast with the lensing efficiency \cite{Kilbinger_2015},

\begin{align}
    \kappa(\bm{X}) &= \frac{3\Omega_{\rm m}H_0^2}{2c^2}\int_0^{\infty}\dd\chi ~q_i(\chi)\frac{\delta_{\rm m}\left(\chi\bm{X}, \chi; z(\chi)\right)}{a(\chi)}\label{eq:kappa-field}\\
    q_i(\chi) &= \int_\chi^\infty \dd\chi' p_i(\chi')\frac{\chi'-\chi}{\chi'},
    \label{eq:lensing-efficiency}
\end{align}
where $\chi$ is the comoving distance, $z(\chi)$ is the redshift at a comoving distance $\chi$ from the observer, $a(\chi)$ is the scale factor, and $\delta_{\rm m}(\bm{r}; z)$ is the matter density contrast at the 3D coordinate $\bm{r}$ and redshift $z$. The lensing efficiency $q_i(\chi)$ is defined for the $i$-th source redshift bin, and characterizes the efficiency of contribution of the matter density field at comoving distance $\chi$ to the convergence field, and depends on the distribution of the source galaxies $p(\chi)$, which is normalized to unity by imposing $\int\dd\chi p(\chi)=1$.

The shear field of weak lensing in Cartesian frame is defined as $\gamma_c(\bm{X})=\gamma_1(\bm{X})+i\gamma_2(\bm{X})$, where $\gamma_1$ and $\gamma_2$ are the shears along the x-axis and an axis rotated 45 degrees from the x-axis, respectively. When the shear is projected onto another reference frame that is rotated by an angle $\zeta$ from the Cartesian frame, the shear field follows the spin-2 transformation
\begin{align}
    \gamma(\bm{X};\zeta)
    &\equiv \gamma_{\rm t}(\bm{X};\zeta)+i\gamma_\times(\bm{X};\zeta)\nonumber\\
    &= -\gamma_{\rm c}(\bm{X}) e^{-2i\zeta}.
\end{align}
where $\gamma_t$ is the tangential component of each galaxy's shear signal, and $\gamma_{\cross}$ its radial component.

We define the Fourier transformation of the shear field as
\begin{align}
    \gamma_{\rm c}(\bm{X}) = \int\frac{\dd^2\bm{\ell}}{(2\pi)^2}\gamma_{\rm c}(\bm{\ell})e^{-i\bm{\ell}\cdot\bm{X}},
\end{align}
and the same for the convergence field. The shear and convergence fields can be related to each other in Fourier space as
\begin{align}
    \gamma_{\rm c}(\bm{\ell}) = \kappa(\bm{\ell})e^{2i\beta}.
    \label{eq:shear-convergence-relation}
\end{align}
Here $\beta$ is the polar angle of the Fourier mode $\bm{\ell}$.

\subsection{Second and third order shear statistics}

To capture the Gaussian information of the shear field, we traditionally rely on two-point statistics. In Fourier space, we write the power spectrum of the convergence field as
\begin{equation}
  \langle\kappa(\bm{\ell}_1)\kappa(\bm{\ell}_2)\rangle 
    = (2\pi)^2\delta^{\rm D}(\bm{\ell}_1+\bm{\ell}_2)P_\kappa(\ell_1).
\label{eq:powerspec}
\end{equation}
The relation between $P_{\kappa}(\ell)$ and the matter power spectrum is given by integrating the latter over redshift with the lensing kernel. For redshift bins $i$ and $j$, we make use of the Fourier space Limber approximation \cite{Kaiser_1992} and write
\begin{equation}
P_{\kappa}(\ell) = \frac{9\Omega_{\rm m}^2H_0^4}{4c^4}\int_{0}^{\infty}\dd\chi\frac{q_i(\chi)q_j(\chi)}{a^2(\chi)}P_{\delta}\left(\frac{\ell}{\chi},z(\chi)\right),
\end{equation}

Analogously, we can extract more information if we move beyond the Gaussian features of the field and consider its three-point statistics. In this way, we use the convergence bispectrum
\begin{align}
  \langle\kappa(\bm{\ell}_1)\kappa(\bm{\ell}_2)\kappa(\bm{\ell}_3)\rangle 
    =& (2\pi)^2\delta^{\rm D}\left(\sum\nolimits_{i=1}^3\bm{\ell}_i\right)\nonumber\\
    &\times B_\kappa(\ell_1, \ell_2, \ell_3).
\end{align}
The relation between the matter bispectrum and the convergence bispectrum can also be given under the Limber approximation \cite{Buchalter_2000}. We have
\begin{align}
\begin{split}
    B_\kappa(\ell_1, \ell_2, \ell_3) &= \frac{27\Omega_{\rm m}^3H_0^6}{8c^6}\int_{0}^{\infty}\dd\chi\frac{q_i(\chi)q_j(\chi)q_k(\chi)}{a(\chi)^3\chi}\\
    &\hspace{2em}\times B_{\delta}\left(\frac{\ell_1}{\chi},\frac{\ell_2}{\chi},\frac{\ell_3}{\chi},z(\chi)\right).
\end{split}
\end{align}

We theoretically model the matter power spectrum and the matter bispectrum in order to have a starting point to compute our $n$-point shear statistics. The non-linear power spectrum is computed through the revised Halofit fitting formula, devised by \cite{Takahashi_2012}. This expression yields an accuracy of 5\% for scales with $k\leq 1 h \text{Mpc}^{-1}$ and 10\% for scales between 1 and 10$h \text{Mpc}^{-1}$ for all our redshifts of interest.
Similarly, we model the matter bispectrum with the BiHalofit formula \cite{Takahashi.Shirasaki.2019}, which is shown to match the perturbation level calculation on Planck 2015 cosmology to an accuracy of $10\%$ on scales with $k<3h\text{Mpc}^{-1}$. Our implementation of the BiHalofit model is described in Appendix~\ref{sec:bihalofit}. 

We can also write, in real space, the shear two-point and three-point correlation functions in terms of the power spectrum and bispectrum. The fact that the shear field is spin-2 gives us three distinct two-point functions by correlating $\langle\gamma_t\gamma_t\rangle$, $\langle\gamma_t\gamma_{\cross}\rangle$, and $\langle\gamma_{\cross}\gamma_{\cross}\rangle$. The correlation between the tangential and radial components is zero for a universe with parity symmetry, leaving us with two independent functions \cite{Kilbinger_2015}. We can combine the tangential and radial components and write

\begin{align}
    \xi_{+}(\theta) \equiv \langle\gamma_t\gamma_t\rangle + \langle\gamma_{\cross}\gamma_{\cross}\rangle , \\
    \xi_{-}(\theta) \equiv \langle\gamma_t\gamma_t\rangle - \langle\gamma_{\cross}\gamma_{\cross}\rangle.
\end{align}

To express the relation between these functions and the convergence power spectrum, we decompose the convergence field into E and B modes. We have now two separate fields $\kappa_E$ and $\kappa_B$ given by

\begin{align}
   \nabla^2 \kappa_E = \boldsymbol{\nabla} (\boldsymbol{\nabla} \kappa) , \\
    \nabla^2 \kappa_B = \boldsymbol{\nabla} \times (\boldsymbol{\nabla} \kappa).
\end{align}

The convergence power spectrum can be decomposed as well between $P^E_{\kappa}(\ell)$ and $P^B_{\kappa}(\ell)$ by replacing $\kappa(\boldsymbol{\ell})$ for its E and B components in Eq.~\ref{eq:powerspec}. Following \citet{Kilbinger_2015}, this brings us to

\begin{align}
    \xi_{+}(\theta) = \int_0^{\infty} \frac{\ell d\ell}{2\pi} J_0(\ell\theta)[P_{\kappa}^E(\ell)+P_{\kappa}^B(\ell)],\\
    \xi_{-}(\theta) = \int_0^{\infty} \frac{\ell d\ell}{2\pi} J_4(\ell\theta)[P_{\kappa}^E(\ell)-P_{\kappa}^B(\ell)],
\end{align}
where $J_{0/4}(x)$ is the 0th-/4th-order Bessel function of the first kind.

For three-point statistics, the traditional convention is to project the shear of a given point over the line that connects it to a chosen center of the triangle, which can be either the orthocenter or the centroid. From this approach, described in \citet{Schneider.Lombardi.2002}, we write eight components of the 3PCF:
\begin{equation}
\gamma_{ijk} = \langle\gamma_i(\bm{X}_1)\gamma_j(\bm{X}_2)\gamma_k(\bm{X}_3)\rangle.
\end{equation}
The so-called natural components of cosmic shear, which are invariant combinations of the above quantities, are defined by \citet{Schneider.Lombardi.2002}. Its real and imaginary parts are:
\begin{equation}
\begin{split}
\textbf{Re}(\Gamma^0) = \gamma_{ttt} - \gamma_{t\times\times} - \gamma_{\times t \times} - \gamma_{\times\times t} \\
\textbf{Im}(\Gamma^0) = \gamma_{tt\times} + \gamma_{t\times t} + \gamma_{\times t t} - \gamma_{\times\times \times}
\end{split}
\end{equation}
\begin{equation}
\begin{split}
\textbf{Re}(\Gamma^1) = \gamma_{ttt} - \gamma_{t\times\times} + \gamma_{\times t \times} + \gamma_{\times\times t} \\
\textbf{Im}(\Gamma^1) = \gamma_{tt\times} + \gamma_{t\times t} - \gamma_{\times t t} + \gamma_{\times\times \times}
\end{split}
\end{equation}
\begin{equation}
\begin{split}
\textbf{Re}(\Gamma^2) = \gamma_{ttt} + \gamma_{t\times\times} -\gamma_{\times t \times} + \gamma_{\times\times t} \\
\textbf{Im}(\Gamma^2) = \gamma_{tt\times} - \gamma_{t\times t} + \gamma_{\times t t} + \gamma_{\times\times \times}
\end{split}
\end{equation}
\begin{equation}
\begin{split}
\textbf{Re}(\Gamma^3) = \gamma_{ttt} + \gamma_{t\times\times} +\gamma_{\times t \times} - \gamma_{\times\times t} \\
\textbf{Im}(\Gamma^3) = - \gamma_{tt\times} + \gamma_{t\times t} + \gamma_{\times t t} + \gamma_{\times\times \times}
\end{split}
\end{equation}

The functions $\Gamma^0$, $\Gamma^1$, $\Gamma^2$ and $\Gamma^3$ can be worked out as functions of the convergence bispectrum. This has been done by \citet{Schneider.Kilbinger.2005}, and is reproduced by \citet{Heydenreich.Schneider.2022} for their three-point function integration code. The expressions involve integrations with $J_2$ and $J_6$ functions, which have a highly oscillatory behavior and thus require caution in direct numerical integration. We have:
\begin{equation}
\begin{split}
    \Gamma^0(x_1,x_2,x_3) = 2\pi\int_0^{\infty}\frac{\ell_1d\ell_1}{(2\pi)^2} \int_0^{\infty}\frac{\ell_2d\ell_2}{(2\pi)^2} \times \\ \int_0^{2\pi}d\varphi B_{\kappa}(\ell_1,\ell_2,\varphi)e^{2i\bar{\beta}} (e^{i(\phi_1-\phi_2-6\alpha_3)}J_6(A_3) + \\e^{i(\phi_2-\phi_3-6\alpha_1)}J_6(A_1)+e^{i(\phi_3-\phi_1-6\alpha_2)}J_6(A_2))
\end{split}
\label{Gamma0eq}
\end{equation}
and
\begin{equation}
\begin{split}
    \Gamma^1(x_1,x_2,x_3) = 2\pi\int_0^{\infty}\frac{\ell_1d\ell_1}{(2\pi)^2} \int_0^{\infty}\frac{\ell_2d\ell_2}{(2\pi)^2} \times \\ \int_0^{2\pi}d\varphi B_{\kappa}(\ell_1,\ell_2,\varphi)e^{2i\bar{\beta}} (e^{i(\phi_1-\phi_2+2\phi_3+2\bar{\beta}-2\varphi-2\alpha_3)}J_2(A_3) + \\e^{i(\phi_3-\phi_2-2\bar{\beta}-2\alpha_1)}J_2(A_1)+e^{i(\phi_3-\phi_1-2\phi_2+2\bar{\beta}+2\varphi-2\alpha_2)}J_2(A_2))
\end{split}
\label{Gamma1eq}
\end{equation}
Here, we express the convergence bispectrum as a function of $\ell_1$, $\ell_2$, and the opening angle between them $\varphi$, for which $\ell_3=(\ell_1^2+\ell^2-2\ell_1\ell_2\cos\varphi)^{1/2}$. We also define the quantity $\bar{\beta}$, which corresponds to the angle between $\ell_3$ and the mean direction between $\ell_1$ and $\ell_2$. 
The values of $\Gamma^2$ and $\Gamma^3$ are obtained through cyclic permutation of the indices on the expression for $\Gamma^1$. The quantities $A_3$ and $\alpha_3$ are defined by the pair
\begin{align}
A_3\sin{\alpha_3} &= (\ell_1x_2-\ell_2x_1)\sin{\frac{\varphi+\phi_3}{2}}, \\
A_3\cos{\alpha_3} &= (\ell_1x_2+\ell_2x_1)\cos{\frac{\varphi+\phi_3}{2}},   
\end{align}
and their counterparts $A_1$ ($A_2$) and $\alpha_1$ ($\alpha_2$) are also given by cyclic permutation of indices.

The requirements for a parameter inference pipeline include being able to generate fast theoretical models to be used by a sampler. Though we can enhance speed with emulation, a faster code is still important to be able to generate a large enough sample for emulator training. The complexity of Eqs.~\ref{Gamma0eq} and \ref{Gamma1eq} calls for a level of precision during integration that leads to an increased runtime. We propose in \citet{sugiyama2024fastmodelingshearthreepoint} a fast algorithm to go from the matter bispectrum to the natural components of the shear three-point correlation function by performing a multipole expansion of the bispectrum. In order to do so, we move away from the conventional projection of the shear, and write our correlation functions relative to the so-called $\times$-projection, defined by \citet{Porth.Schneider.2023} as
\begin{align}
\zeta_1^{\times} &= \frac{\varphi_1+\varphi_2}{2}, \\
\zeta_2^{\times} &= \varphi_1, \\ 
\zeta_3^{\times} &= \varphi_2.
\end{align}
where $\varphi_1$ and $\varphi_2$ correspond to the angles between each of two triangle sides and the x-axis.

Using this convention, we write the expression for the first component of the shear 3PCF:
\begin{align}
\begin{split}
    \Gamma_0^{\times}(\theta_1, \theta_2, \phi) = 
    &-\int\frac{\dd^2\bm{\ell_1}}{(2\pi)^2}\frac{\dd^2\bm{\ell_2}}{(2\pi)^2}
    e^{-i\bm{\ell_1}\cdot\bm{\theta}_1-i\bm{\ell_2}\bm{\theta}_2} \\
    &\times
    B_{\kappa}(\ell_1, \ell_2, \varphi)
    e^{2i\sum_i\beta_i} 
    e^{-3i(\varphi_1+\varphi_2)}. 
    \label{eq:new_gamma0}
\end{split}
\end{align}
By replacing the bispectrum in Eq.~\ref{eq:new_gamma0} by its expansion in Legendre Polynomials, we arrive at a way to compute $\Gamma_0^{\times}$ as a sum of a series of multipole components. The same procedure is done for $\Gamma_1^{\times}$, $\Gamma_2^{\times}$, and $\Gamma_3^{\times}$ and detailed in \citet{sugiyama2024fastmodelingshearthreepoint}.
We show that this method yields a significant speedup relative to brute-force integration. Our \software{fastnc} code is used in our pipeline to model the shear three-point functions.

\subsection{Mass aperture statistics}\label{sec:model-mass-aperture}

The mass aperture is a single measurement of the convergence signal within a circular patch. While the average convergence is, by construction, zero, we can get second-order and third-order shear information by analyzing the moments of the mass aperture. Particularly, the skewness of the mass aperture, given by $\mapcube$, is an efficient way to compress the third-order information that would rather be scattered along multiple triangle configurations on the three-point correlation function. 

The mass aperture statistic can be computed either as an integral of the bispectrum or directly from the natural components of the 3PCF. We opt for the second approach, which allows us to have a theory model consistent with \software{TreeCorr} \citep{Jarvis.Jain.2003} measurements of the mass aperture. In order to guarantee that our model will reproduce the measured statistic, we consistently apply scale cuts and bin averaging on our modeled 3PCF, using the same choices as the ones done for the measurements (detailed in Section~\ref{sec:covariance}). 

With the binned values of the $\Gamma^i$ functions, the mass aperture is found through matrix multiplication. In integral form, the expression was determined by \citet{Jarvis.Jain.2003}, with the $T_0$ and $T_1$ functions defined in their Eqs.~51-52, and the quantities s and $\textbf{t'}$ given in their Eqs.~46-48 as combinations of the vectors $\textbf{q}_i$ that connect each vertex of a triangle configuration to the centroid. Permuting the $\textbf{q}_i$ indices in the $T_1$ expression gives us the functions $T_2$ and $T_3$. The final expression for $\mapcube$ at an aperture radius $\theta$ can then be simplified as

\begin{equation}
\begin{split}
\langle\mathcal{M}_{\rm ap}^3\rangle(\theta) =
\frac{3}{2}\text{Re}\int \frac{sds}{s^2}\int_{s<t'<|t'-s|} \frac{d^2\bm{t}'}{2\pi\theta^2} \\ \times \sum_{i=0,1,2,3}\Gamma^i(s,\bm{t}')T_i\left(\frac{s}{\theta},\frac{\bm{t}'}{\theta}\right)
\end{split}
\end{equation}

In addition to computing the $\mapcube(\theta)$ with the DES-Y3 redshift source redshift bin integration kernels, our pipeline also computes the redshift-dependent $\mapcube(\theta, z)$, which can be transformed into the full mass aperture with a line-of-sight integral. This gives us the advantage of making all the three-point and mass aperture integration independent of the source galaxy redshift distribution. We take advantage of this feature to build our emulator, described in section \ref{sec:nn_emulator}.

The $\mapcube$, as a local measurement, is generally insensitive to partial-sky coverage. On survey data, which commonly have complex masking, the presence of holes and edges can bias a direct measurement of the mass aperture \cite{Secco.Weller.2022}. However, the approach of estimating $\mapcube$ from the three-point correlation function integration removes the mask dependency \cite{Jarvis.Jain.2003}.

It has been shown by \citet{Heydenreich.Schneider.2022} that using $\mapcube$ for cosmological parameter estimation yields results comparable to those obtained by using the most relevant principal components of the full three-point function, and that these first principal components already saturate the available information, making it unnecessary to continue adding the remaining ones. 

\subsection{Modeling observational systematics}\label{sec:systematics-modeling}

An important step towards a robust estimate of the cosmological parameters is to model the effect of observational systematics. If left unmodeled, systematic effects on the data can bring significant biases to the final parameters. Although most of the observational systematics are calibrated at the catalog level in DES-Y3 data, we introduce nuisance parameters in our theoretical expressions in order to marginalize over the potential residual error from these systematics. Our formalism follows the one used by the two-point DES-Y3 analysis \cite{Secco_2022} \cite{Myles_2021} \cite{MacCrann_2021}.

We first include in our model the photometric redshift error, which should account for eventual shifts in the computed redshift distribution of the survey data. To model the residual error in the estimated photometric redshift, we allow the mean redshift of the source distribution to vary by a shift parameter $\Delta z_i$ for $i$-th redshift bin,
\begin{align}
    p_i(z) \rightarrow p_i(z-\Delta z_i).
\end{align}

In terms of shear biases, the DES-Y3 shape catalog uses a self-calibration method to calibrate most of the bias based on the data itself. This is expected to eliminate additive biases on the shear values. However, the blending effect cannot be fully removed, which leaves percent-level systematic uncertainties in the multiplicative bias of the shear. To account for this residual bias in theory, we include one shear multiplicative bias parameter for each source redshift bin, and modulate the 2PCF and 3PCF by
\begin{align}
    &\xi_\pm^{ij} \rightarrow (1+m_i)(1+m_j)\xi_\pm^{ij},\\
    &\Gamma_\mu^{ijk}\rightarrow (1+m_i)(1+m_j)(1+m_k)\Gamma_\mu^{ijk}.
\end{align}
Here $m_i$ is the multiplicative bias for $i$-th source redshift bin, and is marginalized over in the parameter inference process with a proper prior following the Bayesian framework discussed in Section~\ref{sec:method-param-inference}.

Modeling the point spread function (PSF) in the observed image is another important aspect of shear estimation. Inaccuracies in the PSF model over the observed sky are known to introduce two additive bias terms in shear estimation: residual and leakage. These biases give rise to additive PSF correlation terms in the $N$PCF. \citet{Gatti.Wilkinson.2020} and \citet{Secco.Weller.2022} measured the size of the PSF correction terms for the 2PCF and 3PCF, respectively, and concluded that they are subdominant. Therefore, we do not include the PSF terms in our model.

\subsection{Intrinsic alignment}
\label{sec:ialignment}
The measured signal from galaxy ellipticity catalogs is comprised of both the weak lensing signal and the intrinsic alignment (IA) signal. A careful consideration of the latter is important in order to get unbiased constraints from a cosmic shear analysis. Additionally, a consistent modeling of IA in two-point and three-point functions can be responsible, via self-calibration, for a significant gain in information on the cosmological parameters \cite{Pyne.Joachimi.2021}. The non-linear alignment (NLA) model is based on the assumption that the intrinsic galaxy shapes are linearly aligned with the tidal field \cite{Bridle_2007}. The NLA model can be easily included in the two-point and three-point function computations by replacing the lensing efficiency as follows \cite{Gatti.collaboration.2020} \cite{Krause.Weller.2017}.

\begin{align}
    q_i(\chi) \rightarrow q_i(\chi) + f_{\rm IA}\left(z(\chi)\right) p_i(\chi)\frac{\dd z}{\dd \chi}.
    \label{align1}
\end{align}
Here $f_{\rm IA}(z)$ is the redshift dependence of the relation between the intrinsic galaxy alignment and the tidal field,
\begin{align}
    f_{\rm IA}(z) = - A_{\rm IA} \left(\frac{1+z}{1+z_0}\right)^{\alpha_{\rm IA}} \frac{c_1 \rho_{\rm crit}\Omega_{\rm m,0}}{D(z)},
    \label{eq:f-IA}
\end{align}
where $z_0=0.62$ is the pivot redshift, $c_1\rho_{\rm crit}=0.0134$ is a conventional constant value, $D(z)$ is the growth function normalized to unity at $z=0$, and $A_{\rm IA}$ and $\alpha_{\rm IA}$ are the NLA model parameters. 

In the context of two-point analyses, intrinsic alignment models with higher complexity have also been explored. \citet{Blazek.2019} introduce the tidal alignment and tidal torquing (TATT) model, motivated by the need to account for the behavior of spiral galaxies which do not necessarily follow the NLA assumption. Instead of performing a population split, they take a perturbative approach, introducing two new free parameters to probe the amplitude and redshift dependence of the tidal torquing alignment component.

In terms of statistical model selection, \citet{Secco_2022} find a preference for NLA over TATT in modeling DES-Y3 data. Analyses from other surveys have also found no definite need to move from NLA towards models with more parameters. \citet{Dalal.Wang.2023} use HSC Y3 data and determine the shift in $S_8$ when changing between IA models to be insignificant. In a joint DES+KiDS analysis, \citet{joint.des.kids} find a larger shift in $S_8$ but it is partially attributed to the fact that TATT allows their pipeline to explore  lower $S_8$ values in a regime with high tidal torquing alignment, skewing the posterior mean. This is not, therefore, a preference  for any particular model.

For higher-order statistics, it is important to maintain  consistent IA modeling between second and third order correlations. \citet{Pyne.Joachimi.2022} determine that such consistency is possible within the NLA framework, which allows us to use the same $A_1$ and $\alpha_1$ parameters for our two-point and three-point functions. Given these considerations, we opt for model simplicity and implement NLA in our pipeline, through Eqs.~\ref{align1} and \ref{eq:f-IA}.

\subsection{Emulator for redshift-dependent mass aperture}\label{sec:nn_emulator}
We build a neural network emulator for the mass aperture statistic, in order to speed up our cosmological inference pipeline. We use the functions provided by the \software{CosmoPower} framework \cite{Mancini_2022} to create and train the network. Our model inputs are the five cosmological parameters $\Omega_{\rm m}$, $S_8$, $h_0$, $\Omega_b$, and $n_s$. The sum of the neutrino masses $m_{\nu}$ is not included as an emulator input because its dependence is not captured by the BiHalofit modeling, which is calibrated on simulations with fixed $m_{\nu}=0.06$eV. Nonetheless, we do not expect this to yield any biases due to current lensing data lacking sensitivity to constrain neutrino masses. 

For each set of cosmological parameters, the model is trained to compute the redshift-dependent (matter field) $\mapcube(\theta_i, z)$ for $\theta_i = {7', 14', 25', 40'}$ and for a pre-defined set of 36 $z$ values, giving us an output of length 144 (see Appendix \ref{sec:appendix-emu}). We verify that using this set of z values for line-of-sight integration gives us a precision of better than 2\% on $\mapcube(\theta)$ relative to a fiducial theoretical model computed with 30 log bins from $z=10^{-4}$ to $z=0.1$ and 100 linear bins from $z=0.1$ to $z=3$. Generating a z-dependent output for $\mapcube$ allows us to have an emulator that is independent of the source redshift distributions and of the intrinsic alignment kernel.

\begin{figure}
    \centering
    \includegraphics[width=\linewidth]{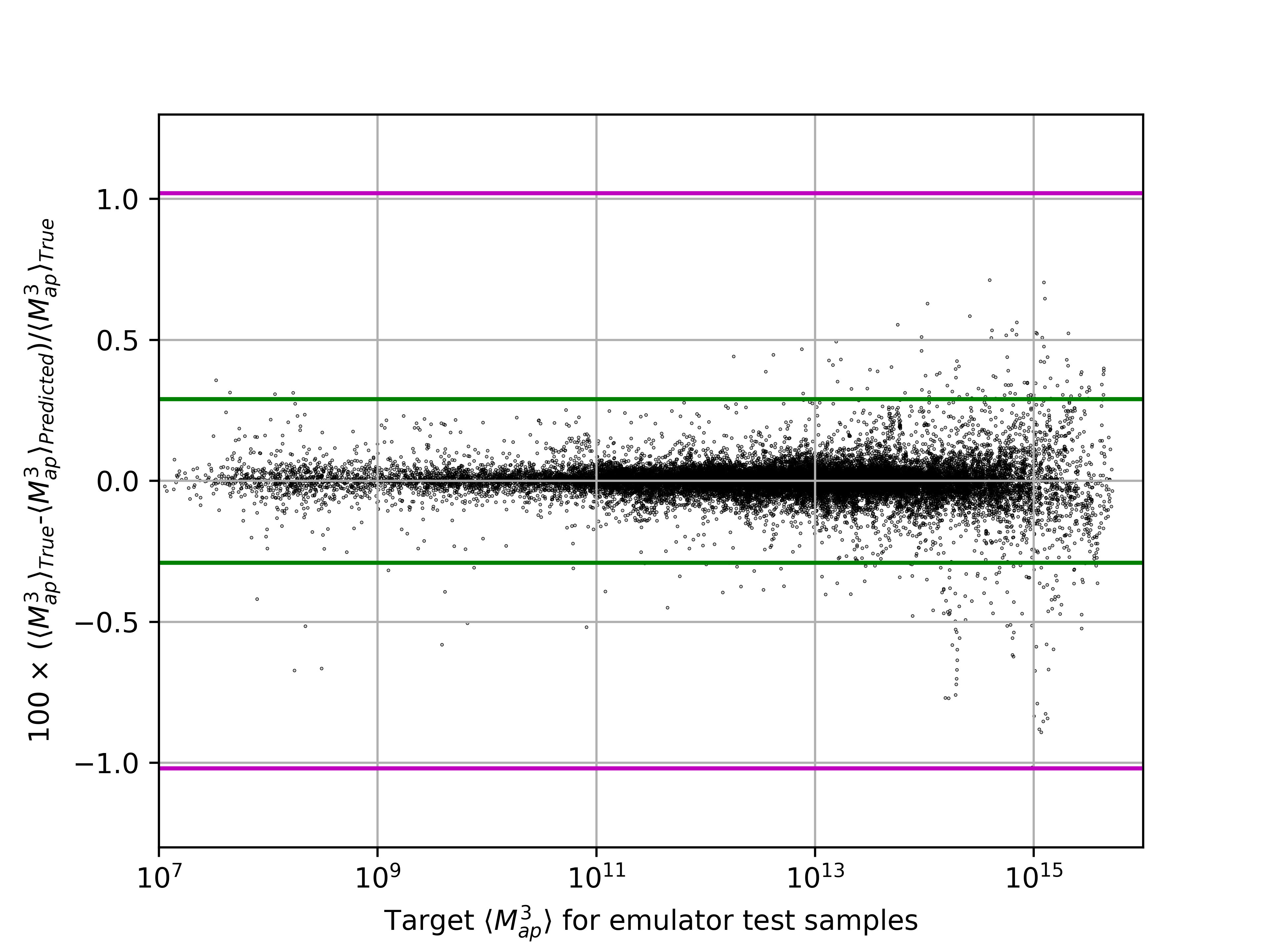}
    \caption{Dispersion of mass aperture emulator predictions. On the y-axis we show the percentage error of the network predictions relative to their corresponding target values. The green lines mark the interval between -0.29\% and 0.29\%, and the magenta between -1.02\% and 1.02\%. A total of $99\%$ of the samples lie between the green lines, and $100\%$ of them between the magenta lines, showing that the emulator scatter is a negligible source of error for our $\mapcube$ model.}
    \label{fig:nn}
\end{figure}

To train and test our model, we generate a set of 1500 samples, drawn as a Sobol Sequence on the five-dimensional parameter space. The range of values used for each parameter is found in Table \ref{table:sobol}. We discard the samples on which the $\mapcube$ computation fails (due to limitations of the Boltzmann solver and of BiHalofit modeling on outlier cosmologies). We divide our remaining samples into one subset for training and validation (with a split of 75\%/25\% and total size of 1300), and a second set of size 171 for testing purposes. We investigate the set of 144 $\times$ 171 outputs for the testing set and find that the network error is below 0.29\% for 99\% of the samples, and below 1.02\% for 100\% the samples. The percentual error on the testing samples is shown in Figure~\ref{fig:nn}.

\begin{table}[h]
    \centering
    \caption{Range of values used to sample our parameter space for emulator training and testing. Values were chosen in order to avoid at most the outlier cosmologies that can break down the $P(k)$ or the $B(k_1,k_2,k_3)$ computation. Our main emulator does not sample the dark energy equation of state $w_0$, leaving it fixed. For our $w$CDM analysis, we build a second emulator and sample it over $w_0$.}
    \label{table:sobol}
    \begin{tabular}{||c|c|c||}
    \hline
    Parameter & min & max \\
    \hline
    \hline
    $\Omega_{\rm m}$ & 0.12 & 0.5 \\ 
    $S_8$ & 0.65 & 0.9 \\
    $h_0$ & 0.55 & 0.91 \\
    $\Omega_b$ & 0.03 & 0.07 \\
    $n_s$ & 0.87 & 1.07 \\
    $w_0$ & -2 & -0.5 \\
    \hline
    \end{tabular}
    \vspace{0.5cm}
\end{table}

In order to use the network predictions on our cosmological inference pipeline, we write the mass aperture $\mapcube(\theta)_{ijk}$ for each redshift bin combination $(i,j,k)$:

\begin{equation}
\mapcube(\theta)_{ijk} = \int \frac{d\chi}{\chi}\frac{q_i(\chi)q_j(\chi)q_k(\chi)}{a(\chi)^3} \mapcube(\theta, z(\chi)) 
\end{equation}

Each integration kernel $q(\chi)$ includes both the lensing kernel and the intrinsic alignment NLA kernel for the selected redshift bin, as described in section \ref{sec:ialignment}.

During integration, the errors from the network output are suppressed. The discrepancy between target and predicted $\mapcube$ for all redshift bin combinations is of the order of 0.01\%, being an insignificant contribution to our total model error budget.

We also developed one additional emulator model to allow for $w$CDM cosmologies. We generated 2000 Sobol samples with the same ranges for $\Omega_{\rm m}$, $S_8$, $h_0$, $\Omega_b$, and $n_s$ as in our fiducial model. We varied the dark energy equation of state $w_0$ parameter between -2 and -0.5. On this new set of samples, we selected 1700 for training and validation (with a 75\%/25\% split), and left the remaining 262 successful ones for testing.
By analyzing the testing samples, we certify that the error is also not significant, remaining below 0.7\% for 99\% of the samples.

By introducing the emulator, the computation speed has significantly improved, reducing the computational time from approximately 40 seconds using \software{fastnc} to just 0.03 seconds with the emulator -- an acceleration by a factor of $\mathcal{O}(10^3)$. This remarkable speed-up makes it feasible for posterior Monte Carlo sampling, enabling efficient exploration of high-dimensional parameter spaces while substantially reducing computational costs.

\section{Covariance Estimation}\label{sec:covariance}

\begin{figure*}
    \centering
    \makebox[\textwidth][c]{%
        \includegraphics[width=1.23\textwidth]{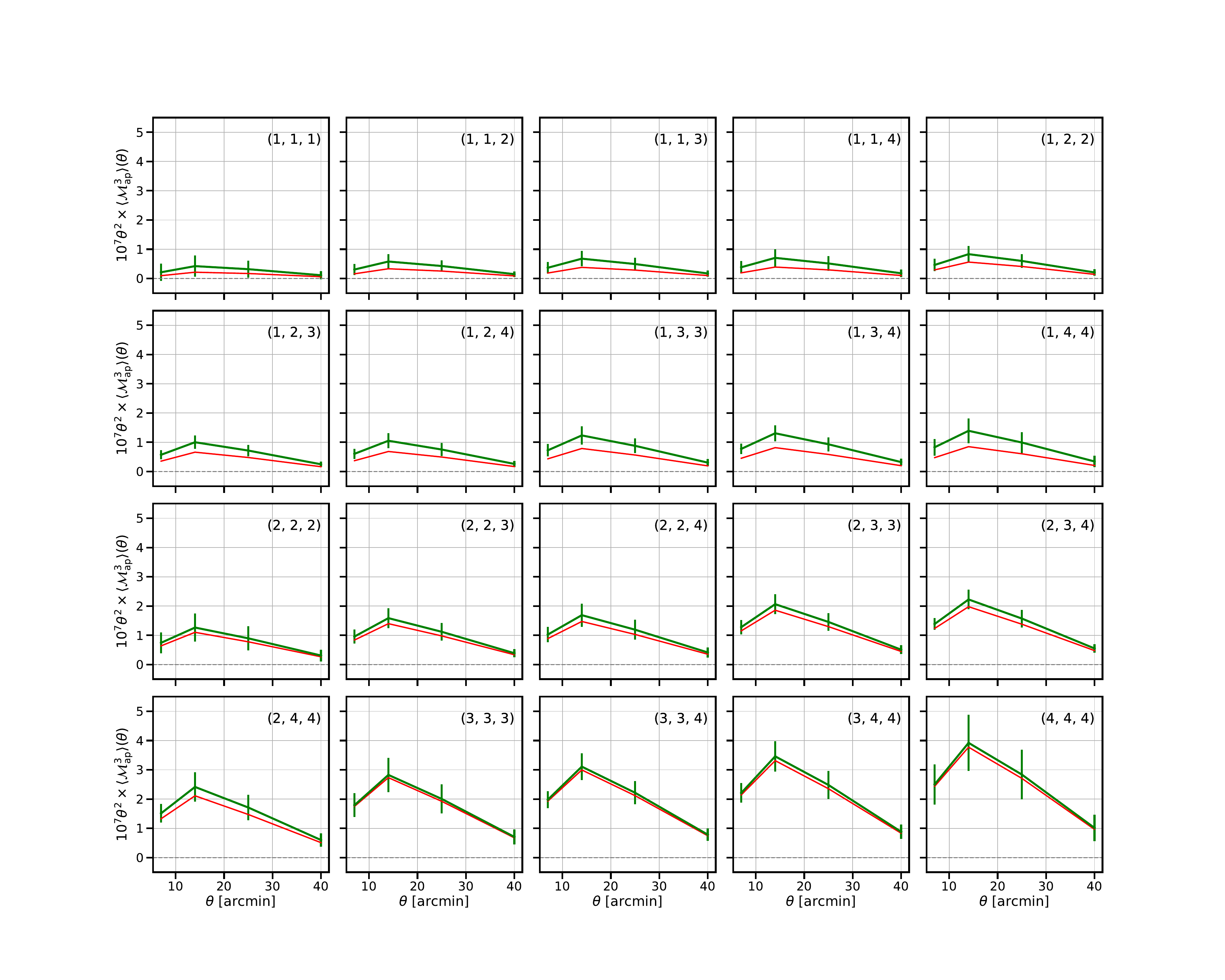}
    }
    \caption{Mass aperture statistic data vector for simulated analysis. Panels show the aperture mass statistic as a function of filter radii $\theta$ for different redshift-bin combinations $(i,j,k)$ indicated on the upper right corner of each panel. The green line indicates our synthetic data vector computed from our pipeline at \software{CosmoGridV1} cosmology. The error bars are estimated from the measured covariance. We also include a red line indicating a theoretical calculation with an artificially increased intrinsic alignment signal to show the effect of this systematic on third-order statistics. We use $A_1=1.0$ and $\alpha_1=0.5$, and find that this is  more relevant for the lower redshift bins.}
    \label{fig:meas}
\end{figure*}

\begin{figure}
    \centering
    \includegraphics[width=\linewidth]{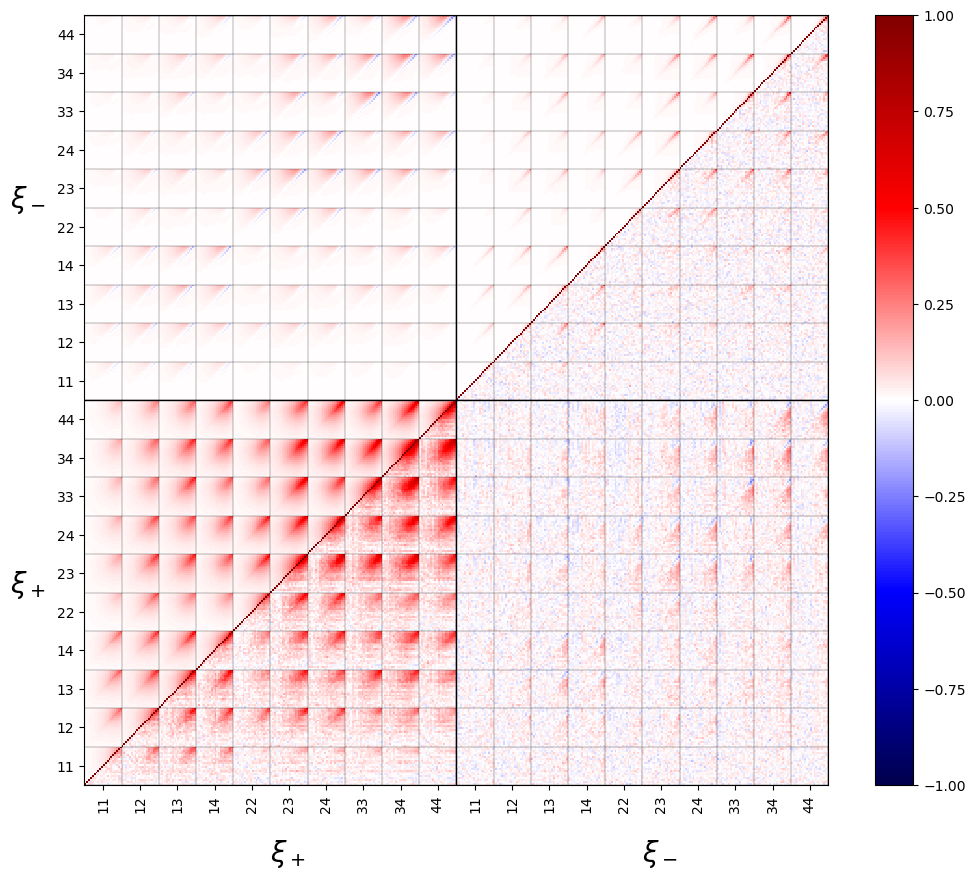}
    \caption{Covariance of second order shear correlation functions. The upper triangle is the analytic result, and the lower triangle is measured from 796 \software{CosmoGridV1} simulations. The covariances are consistent, having sufficiently close amplitude and structure.}
    \label{fig:cov2pt}
\end{figure}

\begin{figure}
    \centering
    \includegraphics[width=\linewidth]{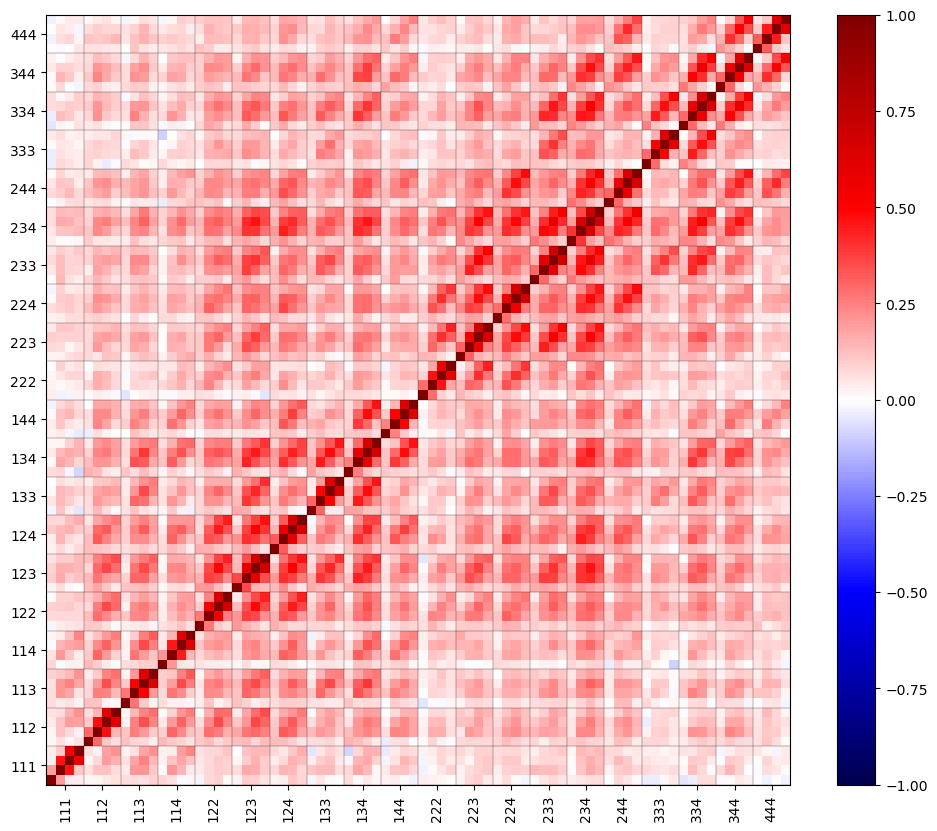}
    \caption{Covariance of the aperture mass statistics estimated from 796 \software{CosmoGridV1} simulations. The number indicated by the ticks on x- and y-axes are the triplets of the redshift bins. The off-diagonal structure amounts mostly to cross-correlations between different aperture filters for the same or nearby redshift bin combinations. This is expected, given that there is overlap between the 3PCF contributions to the $\mapcube$ integral at different filters. We check the convergence of this covariance in Appendix~\ref{sec:appendix-convergence}}
    \label{fig:cov}
\end{figure}

\begin{figure}
    \centering
    \includegraphics[width=\linewidth]{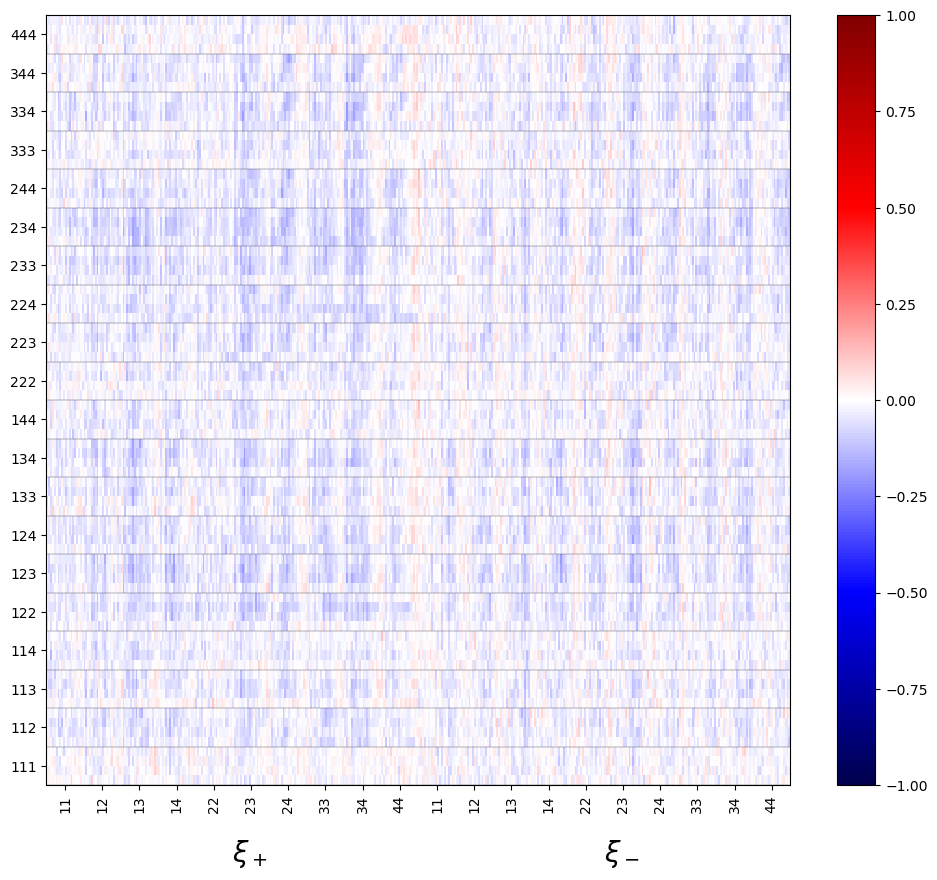}
    \caption{Cross-covariance between the two-point functions and the aperture mass statistics estimated from 796 \software{CosmoGridV1} simulations. The numbers indicated by the ticks on x/y axes are the pairs/triplets of the redshift bins. The lack of noticeable cross-correlation even for  the same redshifts is due to the S/N discrepancy between the statistics. 
    }
    \label{fig:crosscov}
\end{figure}

A joint analysis of second- and third-order shear statistics requires modeling the covariance matrix of the joint data vector. In this paper, we use simulations to estimate the data covariance for $\xi_\pm$ and $\mapcube$. The \software{CosmoGridV1} suite \citep{Kacprzak.Stadel.2022} is a set of lightcone simulations generated with the \software{PkdGrav3} code \cite{pkdgrav3}. It includes runs at a grid of varying cosmologies, as well as runs at a fiducial cosmology. For our covariance estimation, we use their fiducial runs to generate 199 full-sky convergence maps at the DES-Y3 source redshift distributions, which are then each transformed into cosmic shear maps and cut out into four DES-Y3 footprints, yielding us a total of 796 shear maps.  

To add shape noise to our mock data, we randomly rotate the ellipticities of the Y3 shape catalog \cite{Gatti.Wilkinson.2020} at the positions of its galaxies. Next, we add to the randomly rotated $e_1$ and $e_2$ values the mock \software{CosmoGridV1} shear values for each galaxy position. Finally, we 
create an \texttt{NSIDE}$=1024$ \software{HEALPIX} map \cite{Górski_2005} and take, for each pixel, the average over all the final shear values. 

For each realization, we perform a measurement of the 2PCF, $\xi_\pm(\theta)_{ij}$, using \software{TreeCorr} on 20 log-spaced $\theta$ bins ranging from 2.5 to 250 arcminutes, which is the same angular binning as in the DES-Y3 key papers \citep{Abbott_2023}. Here the subscript indicates the pair of redshift bins that are being correlated, and runs over the 10 possible combinations that arise from the four tomographic redshift bin setup.

Similarly, we perform a measurement of third order shear statistics for each realization. We first perform our 3PCF measurements with \software{TreeCorr}, using the Multipole binning scheme. In this scheme, each triangle configuration is described by two side lengths and a set of multipoles of the opening angle. We take as the maximum multipole ${\rm max}_{n}=100$. For the side binning, we take 20 logarithmically spaced bins between $\theta_{\rm min}=0.5'$ and $\theta_{\rm max}=80'$. Next, we convert these measurements into the side-angle-side binning scheme, using 63 linearly spaced bins for the opening angle, which go from 0.016 to 3.126 radians. Finally, these measurements are transformed, using \software{TreeCorr}, into measurements of the skewness of the mass aperture statistic $\mapcube(\theta)_{ijk}$. We use four equal-aperture filters, at $\theta=7$, $14$, $25$, and $40$ arcminutes. As in the case of 2PCF, the subscript indicates the triplet of the redshift bins to correlate, and runs over the 20 possible combinations of a four tomographic redshift bin setup. By restricting ourselves to equal-aperture filters, we avoid an unnecessary increase in our data vector length and retain around 90\% of the full mass aperture information, as investigated by \citet{Burger.Martinet.2023}. 

We concatenate the measured 2PCF and third order mass aperture statistic in a joint data vector, $\bm{d}=[\xi_\pm(\theta)_{ij}, \mapcube(\theta)_{ijk}]$ for each realization, and estimate the covariance by the sample covariance of all the simulation realizations
\begin{align}
    \bm{C} = \frac{1}{N_{\rm real}-1}\sum_{r=1}^{N_{\rm real}} \left[\bm{d}^{r} - \bar{\bm{d}}\right] \left[\bm{d}^{r} - \bar{\bm{d}}\right]^{\rm T},
\end{align}
where $\bm{d}^{\rm r}$ is the measurement of the joint data vector from the $r$-th realization, $\bar{\bm{d}}$ is the mean over all realizations, and $N_{\rm real}=796$.

Through a preliminary comparison of our simulated covariance with jackknife and bootstrap estimates from Y3 data, we verify that our current \software{NSIDE} introduces an inaccuracy in the mass aperture covariance if we do not cut our 3PCF measurements at $\theta \approx 8'$, which is slightly larger than two times the simulation pixel size. Once this low cut is implemented on the two triangle length parameters, there is no need to introduce additional cuts on the maximum multipole parameter. We implement this cut and recompute our $\mapcube$ measurements and joint $\xi_{\pm}$ and $\mapcube$ covariance with the reduced 3PCF data. 

Our mass aperture data vector is presented in Figure \ref{fig:meas}, and was computed with our theory pipeline at \software{CosmoGridV1} cosmology. We use our simulated covariance to compute the total signal-to-noise ratio over all redshift bin combinations, finding it to be S/N=8.9. 
 The highest contributions towards the total S/N ratio can be traced to the redshift bin combinations $(z_1,z_2,z_3)=(3,4,4)$, $(3,3,4)$, and $(2,3,4)$, each yielding individual values of S/N between 6 and 6.3. For this reason, when we explore parameter constraints with subsets of the data, we include a scenario in which we focus on the high-redshift information and add to the auto-correlations the combinations $(3,3,4)$ and $(3,4,4)$.

In order for our likelihood model to bring robust contours on parameter space, we must accurately model three components of our covariance: the two-point correlations, the mass aperture, and the cross-covariance between both (Figures~\ref{fig:cov2pt}, \ref{fig:cov}, and \ref{fig:crosscov}). We confirm the validity of our two-point covariance by comparing it with the analytic model by \cite{Secco_2022, Amon.Weller.2021} (Figure~\ref{fig:cov2pt}). We verify the similarity in structure between the measured and analytic scenarios. The amplitude of both covariances is also consistent, with the average ratio of their diagonal elements being sufficiently close to unity. The effect on the two-point function parameter constraints of switching from an analytic to a simulated covariance is presented on Figure \ref{fig:chain_test_moped}, where we also apply MOPED compression on the simulated covariance chains, as described in Section \ref{sec:method-data-compression}. 

The covariance of our $\mapcube$ data vector is shown in Figure~\ref{fig:cov}. We verify that, within the same redshift bin combinations, there is high cross-correlation between the information from different aperture angles $\theta$. Nonetheless, the cross-correlation over different z-bin combinations is low.

The cross-covariance between two-point and three-point information is also not significant, and there is no structure on the matrix, as can be seen in Figure~\ref{fig:crosscov}. Normalizing the full covariance matrix so that it has a diagonal equal to unity, we find that the largest value for the cross-covariance equals 0.13. We note that this absence of correlation extends even to data vector entries that probe the same redshift bins. Even though it would be reasonable to expect that the (i,i) sections of the $\xi_{\pm}$ data vector would correlate with the (i,i,i) section of the $\mapcube$ vector, at least more than with the other parts of the $\mapcube$ vector, this effect can be suppressed by the S/N discrepancy between the statistics. 

\section{Methodology}\label{sec:methods}
\subsection{Parameter inference}\label{sec:method-param-inference}
\begin{table*}[t]
    \caption{The choices of data vector considered in this paper.}
    \centering
    \begin{tabular}{l|l|c}
    \hline\hline
    label & description & data dim. $N_{\bm{d}}$ \\
    \hline
    {\tt 2pcf+map3}        & $\xi_{\pm}$ and all $\langle\mathcal{M}_{\rm ap}^3\rangle$       & 307 \\
    {\tt 2pcf+map3(auto)} & $\xi_{\pm}$ and auto-$z$ ($z_1=z_2=z_3$) of $\langle\mathcal{M}_{\rm ap}^3\rangle$ & 243 \\
    {\tt 2pcf+map3(auto+334+344)} & adding $(z_1,z_2,z_3)=(3,4,4)$ and $(3,4,4)$ of $\langle\mathcal{M}_{\rm ap}^3\rangle$  to {\tt 2pcf+map3(auto)}& 251 \\
    {\tt 2pcf}             &  only 2pcf & 227 \\
    {\tt map3} & $\langle\mathcal{M}_{\rm ap}^3\rangle$       & 80 \\
    \hline
    {\tt 2pcf[MOPED]+map3}        & MOPED compression of {\tt 2pcf} and all {\tt map3} (This is the fiducial) & 96 \\
    {\tt 2pcf[MOPED]+map3(auto)} & MOPED compression of {\tt 2pcf} and {\tt map3 (auto)} & 32 \\
    {\tt 2pcf[MOPED]+map3(auto+334+344)} & MOPED compression of {\tt 2pcf} and {\tt map3 (auto+334+344)} & 40 \\
    {\tt 2pcf[MOPED]}             & MOPED compression of {\tt 2pcf} & 16 \\
    \hline\hline
    \end{tabular}
    \label{tab:data-vector-choice}
\end{table*}

\begin{table}[t]
    \caption{List of the model parameters and their priors. The first section summarizes the cosmological parameters, and the following sections are nuisance parameters for the observational and astrophysical systematics: the residual photometric redshift uncertainty, the shear multiplicative bias, and the intrinsic alignment of the source galaxies. We employ Gaussian priors $\mathcal{N}$ for parameters of observational systematics (instead of uniform priors $\mathcal{U}$ which are used for cosmological and astrophysical parameters) as these are determined by the  calibration pipelines \citep{Secco_2022}.}
    \centering
    \setlength{\tabcolsep}{15pt}
    \begin{ruledtabular}
    \begin{center}
    \begin{tabular}{ll}
    Parameter & Prior \\ \hline
    \multicolumn{2}{l}{\hspace{-1em}\bf Cosmological parameters}\\
    $\Omega_{\rm m}$        & ${\cal U}(0.1, 0.5)$\\
    $S_8$                   & ${\cal U}(0.7, 0.84)$\\
    $h$                     & ${\cal U}(0.55, 0.91)$\\
    $\Omega_{\rm b}$        & ${\cal U}(0.03,0.07)$\\
    $n_{\rm s}$             & ${\cal U}(0.87, 1.07)$\\ 
    $m_{\nu}~[{\rm eV}]$    & ${\cal U}(0.06, 0.6)$ \\
    $w_0$                     & ${\cal U}(-2.0, -0.3333)$\\
    \multicolumn{2}{l}{\hspace{-1em}\bf Photo-$z$ errors}\\
    $\Delta z_{\rm 1}$      & ${\cal N}(-0.0, 0.018)$  \\
    $\Delta z_{\rm 2}$      & ${\cal N}(-0.0, 0.015)$  \\
    $\Delta z_{\rm 3}$      & ${\cal N}(-0.0, 0.011)$  \\
    $\Delta z_{\rm 4}$      & ${\cal N}(-0.0, 0.017)$  \\
    \multicolumn{2}{l}{\hspace{-1em}\bf Multiplicative shear calibration}\\
    $m_1$                   & ${\cal N}(-0.0063, 0.0091)$  \\
    $m_2$                   & ${\cal N}(-0.0198, 0.0078)$  \\
    $m_3$                   & ${\cal N}(-0.0241, 0.0076)$  \\
    $m_4$                   & ${\cal N}(-0.0369, 0.0076)$  \\
    \multicolumn{2}{l}{\hspace{-1em}\bf Intrinsic Alignment parameters}\\
    $A_{\rm IA}$            & ${\cal U}(-5, 5)$ \\
    $\alpha_{\rm IA}$       & ${\cal U}(-5, 5)$ \\
    \end{tabular}\end{center}
    \end{ruledtabular}
    \label{tab:parameters}
\end{table}


For the parameter inference process, we attempt to fit our theoretical modeling to the data vector through Bayesian analysis,  where the posterior distribution of the model parameters $\bm{p}$ for a given data vector $\bm{d}$ is proportional to the product of the data likelihood and the prior on the model parameters,
\begin{align}
    \mathcal{P}(\bm{p}|\bm{d}) \propto \mathcal{L}(\bm{d}|\bm{p}) \Pi(\bm{p}).
\end{align}

We adopt a non-Gaussian likelihood in order to accurately propagate the variance of our covariance matrix into our final parameter constraints. It has been shown by \citet{Sellentin.Heavens.2015} that, when using a covariance matrix estimated from simulations, a marginalization over the inverse-Wishart distribution of the true underlying covariance will yield a likelihood that follows a modified multivariate t-distribution. An alternative approach is to use a Gaussian likelihood and rescale the inverse covariance by the multiplicative factors introduced by \citet{Hartlap_2006} and \citet{Dodelson.Schneider.2013}. We adopt the likelihood proposed by \citet{Percival.Friedrich.2021}, which follows a t-distribution and is designed to yield Bayesian credible intervals for the model parameters that would match the frequentist-based confidence intervals of the multiplicative factor approach.

Our likelihood is written in terms of the chi-squared difference between the data and the theoretical prediction $\bm{t}(\bm{p})$ for a set of parameters $\bm{p}$:

\begin{equation}
    \ln\mathcal{L}(\bm{d}|\bm{p}) = -\frac{m}{2}\ln{\left(1+\frac{\chi^2}{N_{\text{real}}-1}\right)} + {\rm const}
\end{equation}
\begin{equation}
    \chi^2 = [\bm{d}-\bm{t}(\bm{p})]^{\rm T}\bm{C}^{-1}[\bm{d}-\bm{t}(\bm{p})].
\end{equation}

Here $N_{\text{real}} = 796$ is the number of simulations used to estimate the covariance, and $m$ is a factor given by

\begin{equation}
m = N_{\bm{p}} + 2 + \frac{N_{\text{real}}-1+f_D}{1+f_D},
\end{equation}
with $N_{\bm{p}}$ being the number of model free parameters and $f_D$ the Dodelson-Schneider factor \cite{Dodelson.Schneider.2013}, which is a function of $N_{\text{real}}$, $N_{\bm{p}}$, and the dimension of the data vector $N_{\bm{d}}$:

\begin{equation}
f_D = \frac{(N_{\bm{d}}-N_{\bm{p}})(N_{\text{real}}-N_{\bm{d}}-2)}{(N_{\text{real}}-N_{\bm{d}}-4)(N_{\text{real}}-N_{\bm{d}}-1)}.
\end{equation}

We compute the inverse covariance using the measured covariance matrix from our 796 \software{CosmoGridV1} realizations described in Section~\ref{sec:covariance}. In this paper, we consider several choices for the data vector $\bm{d}$, which are summarized in Table~\ref{tab:data-vector-choice} along with their dimensions $N_{\bm{d}}$. The lower part of the table includes MOPED compression of the data vectors, which is discussed in Section \ref{sec:method-data-compression}.

Our fiducial data vector includes the $\xi_+$ and $\xi_-$ functions for two-point statistics, as well as the $\langle\mathcal{M}_{\rm ap}^3\rangle$ function for three-point correlations. For the $\xi_{\pm}$ part of our data vector, we used 20 log-spaced values of $\theta$ for each source redshift bin combination, with $\theta_{\rm min}=2.5'$ and $\theta_{\rm max}=250'$. This gives us a total length of 200 for $\xi_+$ and 200 for $\xi_-$. To avoid baryonic contamination at the level of the two-point functions, we performed the scale cuts validated by \citet{Secco_2022} for use on DES-Y3 shear data. After applying cuts, our total size for the $\xi_{\pm}$ data vector became equal to 227.

For the mass aperture part of the data vector, we used our full set of 80 data points, which correspond to four aperture values for each of the 20 redshift bin combinations. Our limits of $\theta_{\rm min}=7'$ and $\theta_{\rm max}=40'$ are informed by the tests described in section \ref{sec:method-validation}. Therefore, the length of the fiducial uncompressed two-point plus three-point data vector is 307.

Besides  the full fiducial data vector, we also perform parameter inference with the $\xi_{\pm}$ vector alone, and also with the $\langle\mathcal{M}_{\rm ap}^3\rangle$ alone. Then, we introduce  subsets of  $\langle\mathcal{M}_{\rm ap}^3\rangle$ to the $\xi_{\pm}$ vector, in order to understand which redshift bin combinations add more information. We start with only the redshift auto-correlations, which give us a total data vector size of $4\times4=16$. Next, we add the correlations between bins $(3,3,4)$ and $(3,4,4)$, to include more information from the high-redshift sources, which contribute significantly to the total signal-to-noise. This gives us a data vector of length $6\times4=24$.

For our theory modeling, we considered the set of parameters listed on Table \ref{tab:parameters}. This gives us six cosmological parameters (plus one for our $w$CDM model), and a set of 10 nuisance parameters, over which our results are marginalized. We use priors consistent with the shear two-point analysis of DES-Y3 data described by \citet{Secco_2022}, reproducing their prior ranges for most parameters and being informed by their posteriors for $\Omega_{\rm m}$ and $S_8$, in order to avoid extrapolating the limits of our emulator sampling.

We build our analysis pipeline in the \software{CosmoSIS} package \cite{Zuntz_2015}, which offers an easy interface to the various samplers and calculation modules. For the modeling of the 2PCF theoretical vector, we utilize the existing modules in the \software{cosmosis-standard-library}, a supplemental library tailored for cosmology analysis. We newly build the \software{CosmoSIS} modules for third-order shear statistics, using the \software{fastnc} code \cite{sugiyama2024fastmodelingshearthreepoint}.

We sample the posterior using \software{MultiNest} \cite{Feroz.Bridges.2009} through \software{CosmoSIS}. For the \software{MultiNest} hyperparameters, we use \software{nlive}=500, \software{efficiency}=0.3, \software{tolerance}=0.1. Once the pipeline finishes sampling, we check if the chain covers most of the posterior volume by plotting the weight of samples as a function of sample IDs \cite{Lemos.Weller.2022}.

\subsection{Data compression}\label{sec:method-data-compression}

As we have seen in the last section, the dimension of the joint analysis data vector grows up to 307, which is comparable to the number of \software{CosmoGridV1} simulations used for covariance estimate. In this case, the Hartlap factor can be significant, and even with the Hartlap factor, the simulation-based inverse covariance fails to describe the true covariant structure of the data vector elements. In order to solve this difficulty, we reduce the dimension of the data vector using the Massively Optimised Parameter Estimation and Data compression (MOPED) \cite{Heavens.Vianello.2017,Heavens.Lahav.1999} algorithm. In this algorithm, we transform the original data vector to a compressed data vector using a linear transformation,

\begin{align}
    \bm{d} \rightarrow \bm{B}^{\rm T}\cdot\bm{d},
\label{eq:moped-data-vector}
\end{align}

where the transformation matrix $\bm{B}$ is the two-dimensional matrix with shape $(N_{\bm{d}}, N_{\bm{p}})$, being $N_{\bm{p}}$ the dimension of the model parameter vector. Therefore MOPED compresses the original data vector down to the size of the model parameter vector, $N_{\bm{p}}$. The transformation matrix is designed to maximize the Fisher information after compression, and it is shown that the compression conserves the Fisher information. Specifically, the $i$-th element of the data vector is designed to maximize the $i$-th model parameter's Fisher information.

\begin{figure}
    \centering
    \includegraphics[width=\linewidth]{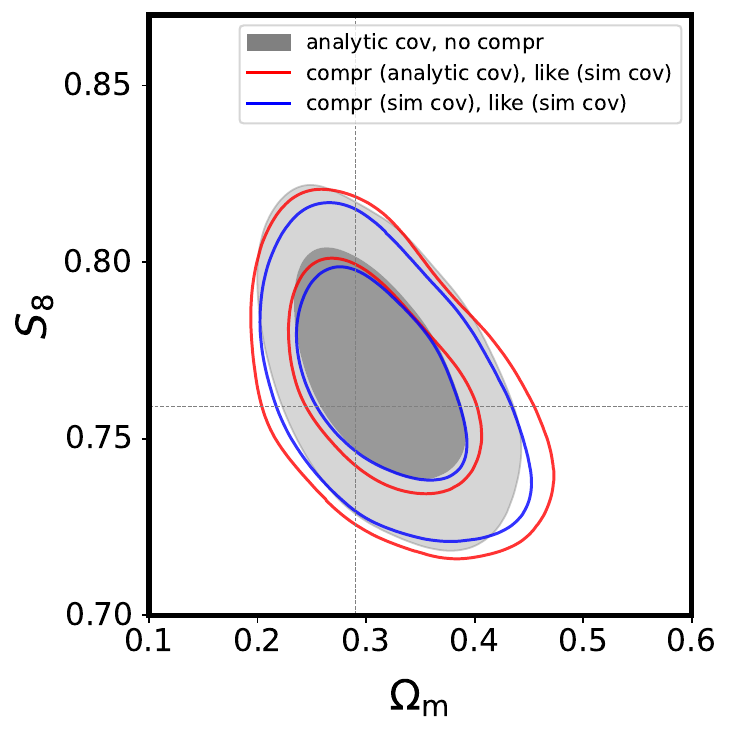}
    \caption{Cosmological constraints with distinct MOPED data compression setups. The gray contour uses the analytic covariance for inference with no compression. The red contour is obtained with the MOPED compressed data vector, where the compression matrix is estimated with the analytic covariance while the simulation-based covariance is used in the likelihood during parameter inference. The blue contour is similar to the red contour, but the simulation-based covariance is used for the estimate of the compression matrix. Here, only the 2PCF data vector is considered. The red contour is slightly larger than the gray due to the difference between the \software{CosmoGridV1} input cosmology and that of the analytic covariance. For our simulated analysis we use the analytic covariance to estimate the compression matrix (red contour). By doing this, we avoid the  overconfidence present in the blue contour.}
    \label{fig:chain_test_moped}
\end{figure}

To obtain the MOPED transformation matrix, we calculate the Fisher matrix, which requires the inverse covariance. One possibility that emerges is to use the rough estimate of the inverse covariance matrix from a small number of simulations. However, we found this not to be an adequate approach. If we use a covariance derived with a simulation number comparable to the dimension of the data vector, the inverse variance of the data elements can be misestimated and the compression matrix can upweight the noisy data vector elements by mistake. This process leads to overconfidence in the parameter inference. 
Figure~\ref{fig:chain_test_moped} shows how much our constraints from the two-point function can be overconfident when we use the simulation-based covariance from a small number of simulations to estimate the compression matrix.

For this reason, we perform a partial data compression only on the 2PCF part of the data vector, for which we have an analytic model of the covariance matrix. We keep our mass aperture data vector uncompressed, and perform MOPED on $\xi_+$ and $\xi_-$, using for compression the analytic model described by \cite{Secco_2022, Amon.Weller.2021}. The performance of the 2PCF compression is shown in Figure~\ref{fig:chain_test_moped}, where we can see that the compression works well to preserve the information content of the 2PCF. The confidence region of the compressed data analysis is slightly larger than that of the original contour due to the larger cosmological amplitude parameter used in \software{CosmoGridV1} simulations relative to that of the analytic covariance.
With the estimated compression matrix for 2PCF, we define the compression matrix for the full data vector by forming the block diagonal matrix.

With the estimated MOPED transformation matrix, we compress the data vector as in Eq.~(\ref{eq:moped-data-vector}), and compute the covariance of the compressed data vector as
\begin{align}
    \bm{C} \rightarrow \bm{B}^{\rm T}\cdot\bm{C}\cdot\bm{B}.
\end{align}

We compute the transformation matrix for all the data vector choices discussed in the last section. The compressed versions of the data vector are summarized in Table~\ref{tab:data-vector-choice}.  

\subsection{Pipeline Validation}\label{sec:method-validation}
\begin{figure}
    \centering
    \includegraphics[width=\linewidth]{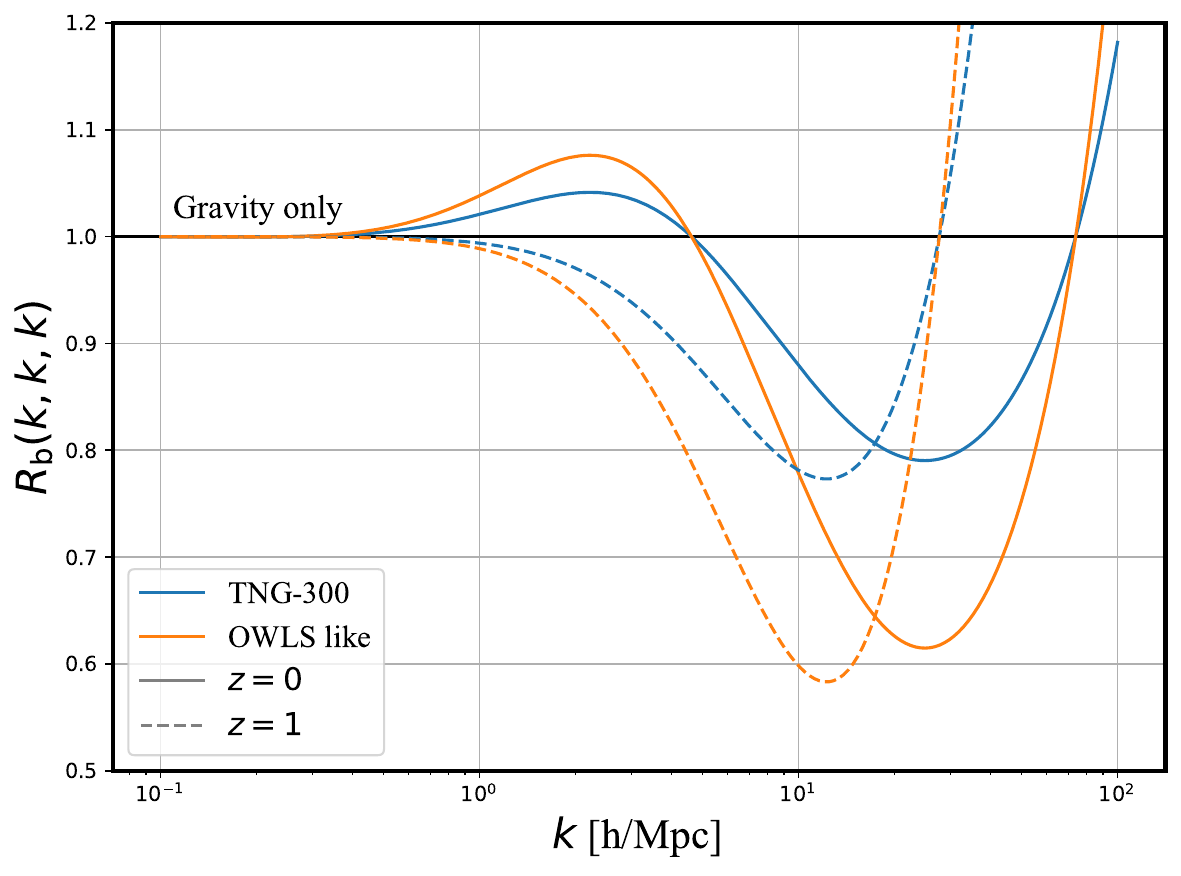}
    \caption{The suppression factors on the matter bispectrum, $R_{\rm b}$, due to baryonic feedback. Here we plot $R_{\rm b}(k_1, k_2, k_3)$ for the equilateral configuration $k_1=k_2=k_3$. The blue lines represent the  TNG-300 simulations, obtained with the fitting formula of  \citet{Takahashi.Shirasaki.2019}. The orange lines represent  the OWLS simulations, obtained by rescaling the blue lines by a factor of $1.5^{3/2}$. The solid and dashed lines are at redshifts $z=0$ and $z=1$, respectively. We use the OWLS-like suppression for pipeline validation.}
    \label{fig:baryon-suppression-bispectrum}
\end{figure}

To perform a robust inference of the cosmological parameters, we need to confirm that our analysis pipeline can recover the underlying true cosmological parameters from data. Thus, we validate our analysis pipeline by performing parameter inference on a simulated data vector at \software{CosmoGridV1} cosmology. We present our results in section \ref{sec:result}.

We also test the validity of our scale cuts against the systematic effect of baryonic feedback on matter clustering. The baryonic effect on matter clustering can be interpreted as the outward displacement of the matter particles in the host halo due to the baryonic physics, e.g. AGN feedback, star formation or gas heating. As a result, the baryonic effect prevents matter clustering at small scales, leading to the suppression in the matter power- and bi- spectra \footnote{Baryonic physics can also cause a bump on the matter bispectrum at an intermediate scale as well as suppression according to \citet{Takahashi.Shirasaki.2019}.}. The physics of these effects is studied in hydrodynamical simulations utilizing sub-grid physical modeling \citep{Schaye.Crain.2014}\citep{Dav.Angles.2019}, but its whole extension is still little understood when compared to gravity-only dynamics, which can lead to uncertain estimates of the resultant suppression.

Because of this difficulty in modeling the baryonic effect on the matter power- and bi- spectra, we do not include baryonic physics in our model, but rather try to remove from the data vector the scales that could be contaminated by the baryonic effect. To this end, we generate a synthetic data vector of the 2PCF and the mass aperture that includes a baryonic effect similar in amplitude to that of the OWLS simulations \cite{Schaye.Vecchia.2010}.
For the 2PCF, we estimate the suppression factor for $P(k)$ from the OWLS simulations, multiply it with the gravity-only non-linear matter power spectrum, and then transform it to the $\xi_{\pm}$ observable. For third-order shear statistics, we start by computing the suppression factor for the matter bispectrum through the fitting formula in Appendix~C of \citet{Takahashi.Shirasaki.2019}. This expression was calibrated from the TNG-300 simulation, which has a baryonic effect smaller in amplitude than that of the OWLS simulations. In order to have a consistent baryonic effect between 2PCF and third-order shear statistics, we rescale the suppression factor accordingly. The parameter $R_{b}$ is defined as the ratio between the matter bispectrum with ($B_b$) and without ($B_{\text{DM}}$) baryons:

\begin{equation}
R_b(k_1,k_2,k_3) \equiv \frac{B_b(k_1,k_2,k_3)}{B_{\text{DM}}(k_1,k_2,k_3)}
\end{equation}

For scales in which the baryonic effect is important, we can approximate the expressions for the matter power- and bi- spectra to the 1-halo regime, in which $P(k) \propto u(k)^2$ and $B(k,k,k) \propto u(k)^3$, with $u(k)$ being the halo profile. Thus, we have

\begin{equation}
\frac{B_b(k,k,k)}{B_{DM}(k,k,k)} \propto \left(\frac{P_b(k)}{P_{\text{DM}}(k)}\right)^{3/2}
\end{equation}

We compare the peaks in the matter power spectra suppression factors from the OWLS and TNG-300 simulations, estimating them to differ by a factor of 1.5. Therefore, we rescale our original $R_b(k_1,k_2,k_3)$ parameter by a factor of $1.5^{3/2}$, so that it has an amplitude consistent with that of OWLS.  The resultant suppression factor is shown in Figure~\ref{fig:baryon-suppression-bispectrum}. Using the suppressed matter bispectrum, we compute the shear 3PCF and mass aperture statistics, and perform a run of our pipeline on the baryon-contaminated data vector. We present our results in the following section, along with those of the full simulated analysis.

\section{Results}\label{sec:result}

In this section, we show the results of our analysis pipeline validation. We measure our level of joint improvement on the $\Omega_{\rm m}$ and $S_8$ constraints through the figure-of-merit (FoM) defined by

\begin{equation}
    \text{FoM} = \frac{1}{\sqrt{\text{Cov}(\Omega_{\rm m}, S_8)}}.
\end{equation}

\begin{figure}
    \centering
    \includegraphics[width=\linewidth]{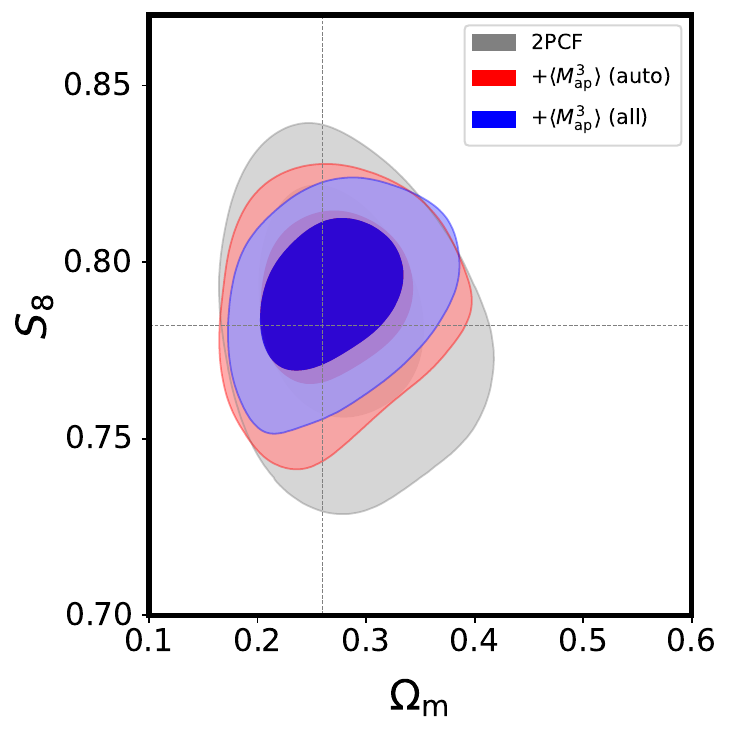}
    \caption{Results of our simulated 2+3pt cosmological analysis. The gray contour is the constraint from $\xi_{\pm}$ alone, while the colored contours are constraints the addition of  the third order mass aperture statistic. The dotted lines indicate the input values of $\Omega_{\rm m}$ and $S_8$. We find an improvement of $83\%$ on the $\Omega_{\rm m}$-$S_8$ figure-of-merit when we add the full mass aperture data vector.}
    \label{fig:chain_test_oms8}
\end{figure}

\begin{figure*}
    \centering
    \includegraphics[width=0.9\linewidth]{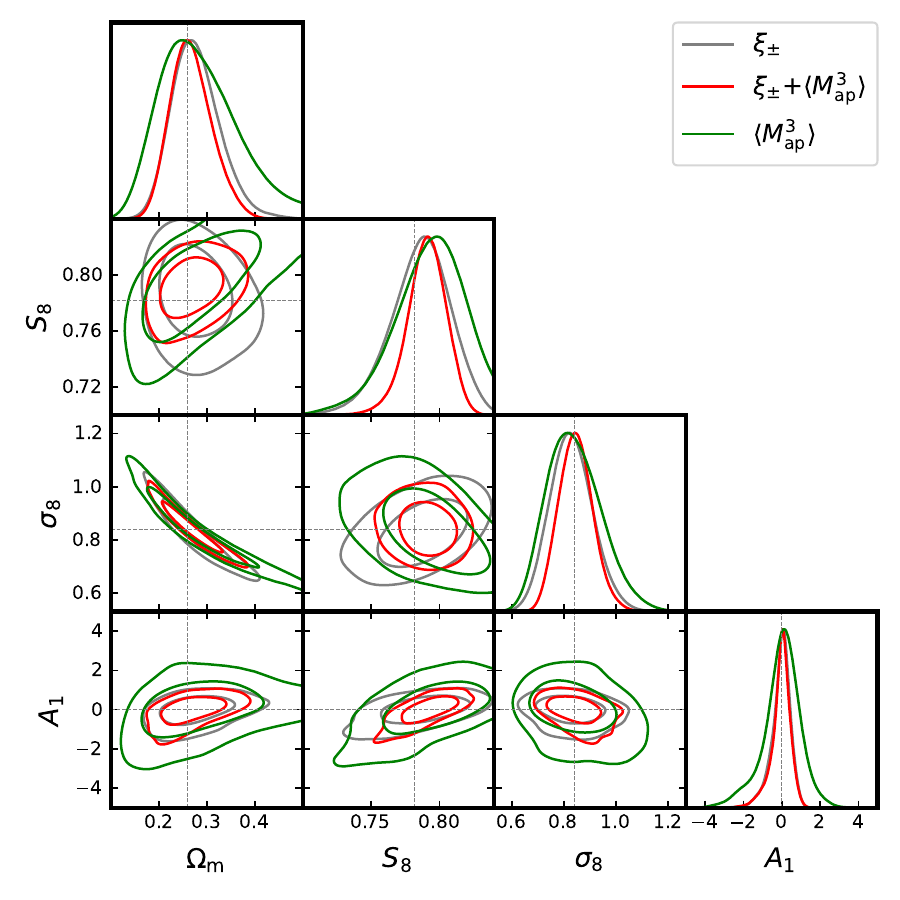}
    \caption{Results of our simulated 2+3pt cosmological analysis. The gray constraint uses 2PCF data alone, the red constraint uses both 2PCF and mass aperture, and the green constraint uses mass aperture data alone. The dotted lines indicate the input values of the parameters. Note the difference in orientation of the $\Omega_{\rm m}$-$S_8$ contours from $\xi_{\pm}$ and $\mapcube$, which contributes to the  improvement in the joint constraint.}
    \label{fig:chain_allparams}
\end{figure*}

Our full run of the pipeline uses synthetic $\xi_{\pm}$ and $\mapcube$ data at the \software{CosmoGridV1} input cosmology. We use our \software{CosmoGridV1} simulated covariance and perform MOPED compression on the $\xi_{\pm}$ data vector before concatenating it with the mass aperture information. 

With $\Lambda$CDM modeling, we find that adding the auto-correlations of the mass aperture increases the FoM on the $\Omega_{\rm m}$-$S_8$ plane by $40\%$, while adding the full mass aperture data improves the 2PCF constraints on $\Omega_{\rm m}$-$S_8$ by $83\%$. The results are shown in Figures~\ref{fig:chain_test_oms8} (focused on $\Omega_{\rm m}$-$S_8$) and \ref{fig:chain_allparams} (which includes the constraint from mass aperture alone). The figure of merit for each considered scenario on $\Omega_{\rm m}$-$S_8$ is presented in Table~\ref{tab:fig_of_merit}. Adding the high signal-to-noise redshift bin combinations (3,3,4) and (3,4,4) of the mass aperture along with its auto-correlations only slightly increased the gain relative to the scenario restricted to the auto-correlations. This shows us that, despite those combinations bringing important contributions to the signal, it is still necessary to combine low and high redshift information to achieve the full constraining power of the mass aperture. Our incremental approach demonstrates the presence of valuable cosmological information at all redshifts. 

We attribute the high FoM gain of $83\%$ to the fact that the individual contours from $\mapcube$ and $\xi_{\pm}$ are nearly orthogonal to each other in the $\Omega_{\rm m}$-$S_8$ plane, inducing a significant reduction of the joint contour, as can be seen in Figure~\ref{fig:chain_allparams}. The contours for all cosmological parameters are presented in Appendix~\ref{sec:allcontours} for both $\Lambda$CDM and wCDM models.

\begin{table}[t]
    \caption{Figure of merit on the $\Omega_{\rm m}$-$S_8$ plane for our different data vector choices. The rightmost column summarizes the gain in the FoM compared to the 2PCF-only analysis.}
    \centering
    \setlength{\tabcolsep}{15pt}
    \begin{ruledtabular}
    \begin{center}
    \begin{tabular}{lcc}
    Data vector & FoM & FoM gain\\ \hline
    \tt{2pcf}        & 930.8 & -- \\
    \tt{2pcf+map3(auto)} & 1299.2 & 40\%\\
    \tt{2pcf+map3(auto+334+344)} & 1408.7 & 52\%\\
    \tt{2pcf+map3(all)} & 1700.8 & 83\%\\
    \hline
    \tt{wCDM-2pcf} & 653.2 & -- \\
    \tt{wCDM-2pcf+map3(all)} & 928.1 & 42\%\\
    \end{tabular}\end{center}
    \end{ruledtabular}
    \label{tab:fig_of_merit}
\end{table}

We verify the consistency between the two-point analysis and our full analysis by comparing the mean of the marginalized parameter posteriors. We also compare those values with the \software{CosmoGridV1} input cosmological parameters. The results are shown in Table~\ref{tab:param_values}.

\begin{table*}[t]
    \caption{Mean of parameter posteriors from 2PCF and 2PCF+$\mapcube$ analysis.}
    \centering
    \setlength{\tabcolsep}{15pt}
    \renewcommand{\arraystretch}{1.5}
    \begin{ruledtabular}
    \begin{tabular}{llllll}
    Parameter & \software{CosmoGridV1} & $\xi_{\pm}$ & $\xi_{\pm}$ + $\mapcube$ & $\xi_{\pm}$ (wCDM) & $\xi_{\pm}$ + $\mapcube$ (wCDM)\\ \hline
    $\Omega_{\rm m}$ & 0.26& $0.277^{+0.049}_{-0.048}$ & $0.268\pm 0.044$  & $0.267^{+0.052}_{-0.051}$  & $0.254^{+0.041}_{-0.042}$ \\
    $S_8$ &0.782& $0.786^{+0.022}_{-0.021}$ & $0.790^{+0.014}_{-0.013}$ & $0.764\pm 0.032$  &  $0.768\pm 0.031$\\
     $w_0$ & -1 & - & -  & $-1.28\pm 0.35$ &$-1.29\pm 0.35$  \\
    \end{tabular}
    \end{ruledtabular}
    \label{tab:param_values}
\end{table*}

\begin{figure*}
    \centering
    \includegraphics[width=\linewidth]{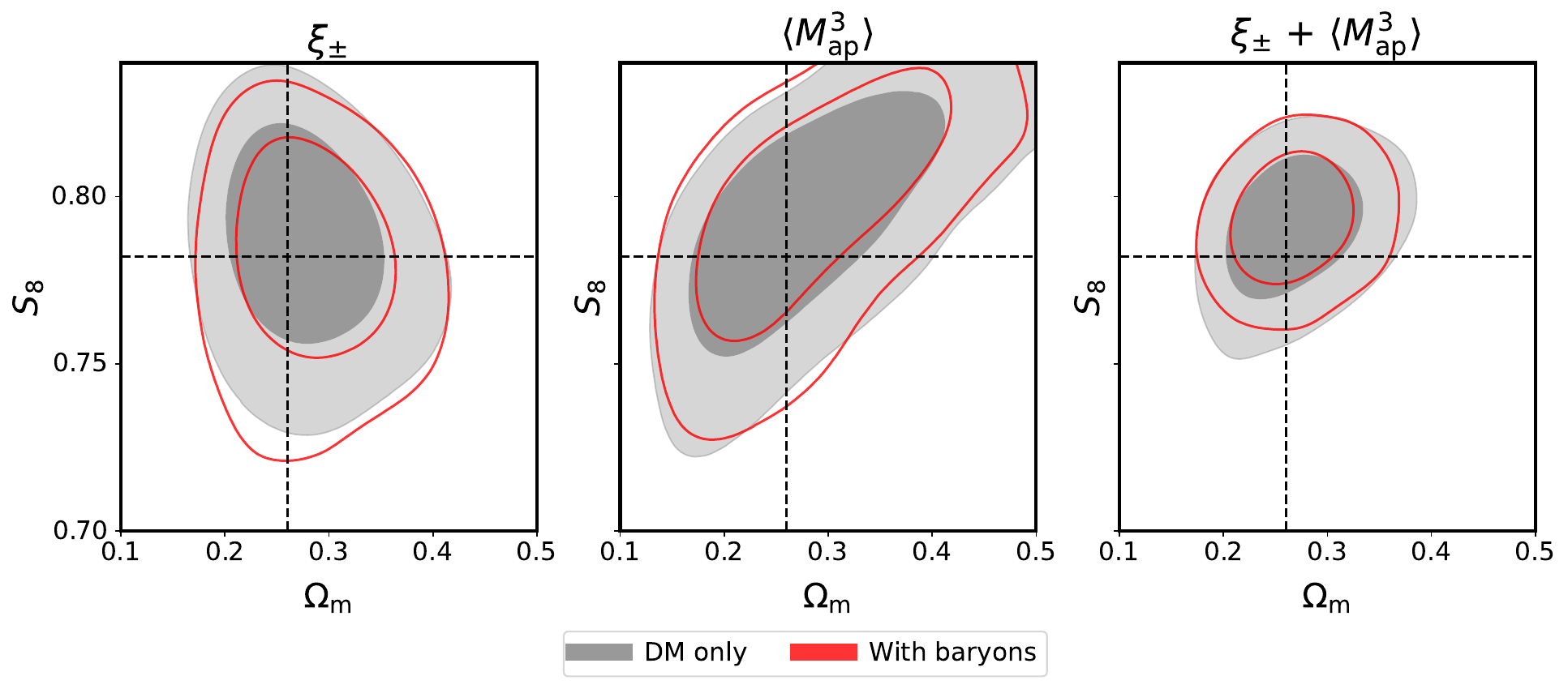}
    \caption{Validation of our analysis pipeline against the effect of baryonic feedback. The gray contours are the results with the dark matter (DM) only data vector, while the red contours use the baryon contaminated data vector. The first panel includes only 2PCF data, the second panel only $\mapcube$ data, and the third panel uses the full data vector. Our baryon-contaminated data vector uses an amplitude similar to that seen on the OWLS simulations (see also Figure~\ref{fig:baryon-suppression-bispectrum} for the baryonic effect on the bispectrum). We find no significant shift in $\Omega_{\rm m}$ and $S_8$ for any of the data sets.}
    \label{fig:chain_test_baryon}
\end{figure*}

To verify the adequacy of our 2PCF and mass aperture scale cuts in removing effects from baryonic physics,
we run our pipeline on our baryon-contaminated data vector which mimics the baryonic effect present in the OWLS simulations (see Section~\ref{sec:method-validation} for the detail). The results are shown in Figure~\ref{fig:chain_test_baryon}. We measure the shift in the mean of the $S_8$ and $\Omega_{\rm m}$ posteriors caused by introducing baryons on the data vector. For the joint analysis, $S_8$ is shifted by 0.05$\sigma$ and $\Omega_{\rm m}$ by 0.0$\sigma$, which is completely insignificant. This indicates that our choice of scale cuts for 2PCF and mass aperture is large enough to cut out the small-scale data vector elements that are more sensitive to the baryonic effect.

We also note that while the presence of baryons slightly moves down the $S_8$ value on the 2PCF analysis, it slightly increases the $S_8$ value on the $\mapcube$-only analysis. The net effect on a joint analysis is that of a smaller shift than in each of the individual analyses. As can be seen in Figure~\ref{fig:baryon-suppression-bispectrum}, there is a regime of baryonic enhancement on the bispectrum for mid-to-low scales. Due to our cuts and filter choices, the enhancement dominates over the smaller-scale suppression regime on the $\mapcube$ integral, leading to a net enhancement of the $\mapcube$ signal.

\begin{figure*}
    \centering
    \includegraphics[width=0.75\linewidth]{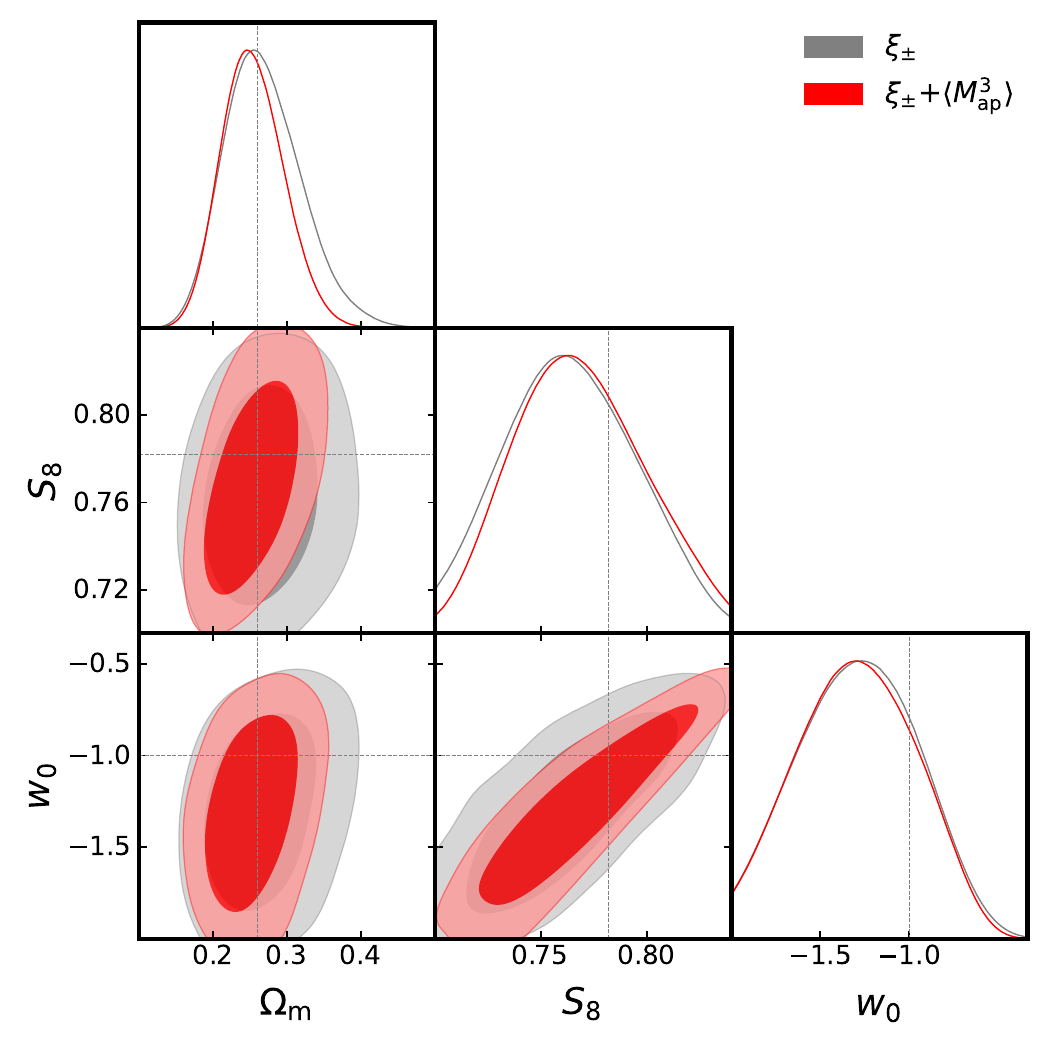}
    \caption{Results of our simulated analysis for the $w$CDM cosmological model. The gray and red contours are from 2PCF-only and 2PCF+$\mapcube$, respectively. The dotted lines are the input values of the parameters. We find an improvement of $36\%$ on the joint $w_0$-$S_8$ constraint.}
    \label{fig:chain_test_oms8_wCDM}
\end{figure*}

Next, we show the improvement of the cosmological parameter constraints in the $w$CDM model. Here we use the same data vector as in our previous $\Lambda$CDM analyses, but we allow the equation-of-state parameter $w_0$ of dark energy to freely vary within the prior range specified in Table~\ref{tab:parameters}. The constraints are shown in Figure~\ref{fig:chain_test_oms8_wCDM}. We find no improvement on the marginalized $w_0$ uncertainty when adding third order shear information to the 2PCF signals. The individual $\Omega_{\rm m}$ gain was found to be of $20\%$, while the $S_8$ gain was $2\%$. In contrast, the marginalized parameter gains in the $\Lambda$CDM scenario were of $10\%$ for $\Omega_{\rm m}$ and $36\%$ for $S_8$, as can be seen in Table~\ref{tab:param_values}. Despite the negligible $S_8$ and $w_0$ individual gains, the improvement in the $w_0$-$S_8$ figure-of-merit was found to be of $36\%$, indicating that mass aperture data helps to break the existing $w_0$-$S_8$ degeneracy. 

Finally, in addition to constraints on the cosmological parameters, we also find a slight improvement on the intrinsic alignment amplitude parameter as detailed in Figure~\ref{fig:chain_test_IA}. With the $\xi_{\pm}$ data alone, we find $A_1=-0.05^{+0.47}_{-0.48}$. When we add the mass aperture information, this constraint moves to $A_1 = -0.05^{+0.48}_{-0.45}$, being tightened by a factor of 3\%. As the amplitude of gravitational lensing is larger for higher source redshift bins due to the larger line-of-sight coverage, the relative contribution of intrinsic alignment is larger at lower redshift bins. This can be seen in Figure~\ref{fig:meas}, where the predicted signals with and without intrinsic alignment are shown.

\section{Conclusions}\label{sec:conclusion}
We have presented results from a pipeline to obtain robust constraints on cosmological parameters using second- and third-order shear statistics. The full configuration and redshift-dependent shear three-point correlations are measured from simulated shear catalogs and compressed for computational efficiency into the third moment of the mass aperture statistic $\mapcube$. We build a theoretical model for $\mapcube$ and integrate it into the \software{CosmoSIS} framework. The methodology developed in our earlier work \cite{sugiyama2024fastmodelingshearthreepoint} enables rapid computation of $\mapcube$ at every point in parameter space. 
Building upon this, we introduce an emulator that further accelerates the computation, reducing the runtime from approximately 40 seconds to just 0.03 seconds -- an improvement by a factor of $\mathcal{O}(10^3)$. This substantial speed-up makes posterior Monte Carlo sampling feasible and significantly enhances computational efficiency.

We compute the  combined likelihood of the two-point functions $\xi_{\pm}$ and of $\mapcube$, including the  covariance matrix from a set of 796 \software{CosmoGridV1} shear maps. We verify the robustness of our pipeline against observational systematics and show that our scale cuts  sufficiently remove contamination from baryonic feedback. We thus complete the preparation of our pipeline for an analysis with DES-Y3 data.

We also determine the improvement in the $\Omega_{\rm m}$-$S_8$ figure-of-merit  of our second- and third-order shear analysis relative to the $\xi_{\pm}$ data vector alone. We obtain a significant gain of $83\%$, which is much higher than the gains reported by \citet{Gatti.Wiseman.2023} in studies of the third moment with DES-Y3. While there are several previous analyses of third-order correlations, \citet{Burger.Martinet.2023} is the closest to our work as they use the full three-point correlations and find a similar gain ($93\%$) for a KiDS simulated analysis with COSEBIs and $\mapcube$. 

We study the complementarity of three-point information in helping to break degeneracies between cosmological and systematic parameters. As noted in early studies of three-point correlations, their dependence on $\sigma_8$ and $\Omega_{\rm m}$ is different from that of two-point correlations \cite[see][for a perturbation theory description]{Bernardeau:1996un}. Our Figure~\ref{fig:chain_allparams} shows that in practice the degeneracy directions differ only subtly, but they do enable stronger constraints on both parameters from a joint analysis. More quantitatively, the two-point correlation function is sensitive to the combination of $S_8\equiv \sigma_8(\Omega_{\rm m}/0.3)^{0.5}$ \footnote{This is the conventional $S_8$ definition, while we found that $\sigma_8(\Omega_{\rm m}/0.3)^{0.53}$ is the parameter combination that the two-point alone analysis in this paper constraints most tightly.}, while we found the three-point is to the combination of $S_8'\equiv \sigma_8(\Omega_{\rm m}/0.3)^{0.43}$.
The full parameter constraints from our joint analysis can be seen in Figures~\ref{fig:lcdm_allparams} and \ref{fig:wcdm_allparams}. 

Interestingly, for our scale cuts on the three-point correlation function and aperture filters between $7'$ and $40'$, the lensing signal is slightly enhanced for $\mapcube$ due to the presence of baryons. Thus, the mass aperture analysis has the opposite bias on the $S_8$ constraint compared to the two-point analysis. The combined effect on a joint analysis is a negligible shift in $S_8$ (below $0.1\sigma$), and thus one of robustness to baryonic feedback relative to that of the individual summary statistics. How generally this holds is an interesting question for future work. 

The gain in the $\Omega_{\rm m}$-$S_8$ FoM that we find with the full three-point correlation raises interesting questions regarding the use of higher order statistics (HOS). In general, various studies have reported complementary information in an array of HOS. In particular, \citet{Gatti.Wiseman.2023} compare the $\Omega_{\rm m}$-$S_8$ FoM gain from third moments, scattering transforms, and wavelet phase harmonics, with each yielding improvements of $15\%$, $38\%$ and $53\%$ relative to second moments alone. The combination of the three statistics yields an improvement of $92\%$ relative to a second moments analysis. The gain in FoM they find with all HOS is very close to the gain we report from three-point correlations alone. While those numbers should not be  compared in detail with ours (in particular because  we use $\xi_{\pm}$ as our second order statistic), they motivate the use of the full three-point function rather than smoothed moments, and of a careful examination of what information remains to be captured by other HOS. We do believe that the three-point function ought to be the first HOS to be included beyond two-point analyses, due to its ease of modeling and interpretation. 

Interestingly, for three-dimensional clustering analyses, recent findings suggest that the power spectrum and bispectrum come close to saturating the information content \cite{Cabass:2023nyo}. Again, there are several differences in the analyses (in particular, the presence of galaxy bias which necessitates more conservative scale cuts), and the question may not be settled yet \cite[see, e.g.][for the field-level inference argument]{Nguyen:2024yth}. 

Finally, we also consider the $w$CDM model, for which we present results for the three key cosmological parameters ($S_8, \Omega_{\rm m}$ and $w_0$). We find an improvement of $36\%$ in the joint $w_0$-$S_8$ constraint. The marginalized uncertainty on  $w_0$ itself shows no improvement \footnote{This contrasts with the $30\%$-$45\%$ improvement  found by \citet{Gong_2023} in their analysis of the integrated shear three-point function. One major difference between the studies is in the lower bounds of the $w_0$ prior. While we impose that $w_0>-2$, \citet{Gong_2023} use $w_0>-3.33$.  This removes a region of parameter space which is well constrained by the combination of second- and third-order statistics.}.

In conclusion, we have shown that using the full three-point correlation functions is a promising way to extract  significant additional cosmological information from lensing surveys. The three-point function extracts non-Gaussian information while staying well above the small scales that are impacted by baryonic feedback. The application to DES-Y3 data will follow in a separate publication. 

\begin{acknowledgments}
We thank Ryuichi Takahashi,  Gary Bernstein and Masahiro Takada for the useful discussions. We also thank Kazuyuki Akitsu for providing references to studies on galaxy clustering.
BJ and RCHG are  partially supported by the US Department of Energy grant DE-SC0007901 and SS is supported by the JSPS Overseas Research Fellowships. Part of this work was supported by the NASA ROSES grant 22-ROMAN11-0011 via a JPL subaward.

Author Contributions: RCHG and SS developed the pipeline, performed the analysis, and wrote the manuscript. BJ served as project advisor and contributed to the manuscript. MJ developed novel \software{TREECORR} functionalities. DA generated the \software{CosmogridV1} mocks used in this work. MG contributed with the two-point analysis setup. DG, ZG, and AH contributed by testing the simulations. AH, GAM and SP served as collaboration internal reviewers, and JLM as collaboration final reader.

Funding for the DES Projects has been provided by the U.S. Department of Energy, the U.S. National Science Foundation, the Ministry of Science and Education of Spain, 
the Science and Technology Facilities Council of the United Kingdom, the Higher Education Funding Council for England, the National Center for Supercomputing 
Applications at the University of Illinois at Urbana-Champaign, the Kavli Institute of Cosmological Physics at the University of Chicago, 
the Center for Cosmology and Astro-Particle Physics at the Ohio State University,
the Mitchell Institute for Fundamental Physics and Astronomy at Texas A\&M University, Financiadora de Estudos e Projetos, 
Funda{\c c}{\~a}o Carlos Chagas Filho de Amparo {\`a} Pesquisa do Estado do Rio de Janeiro, Conselho Nacional de Desenvolvimento Cient{\'i}fico e Tecnol{\'o}gico and 
the Minist{\'e}rio da Ci{\^e}ncia, Tecnologia e Inova{\c c}{\~a}o, the Deutsche Forschungsgemeinschaft and the Collaborating Institutions in the Dark Energy Survey. 

The Collaborating Institutions are Argonne National Laboratory, the University of California at Santa Cruz, the University of Cambridge, Centro de Investigaciones Energ{\'e}ticas, 
Medioambientales y Tecnol{\'o}gicas-Madrid, the University of Chicago, University College London, the DES-Brazil Consortium, the University of Edinburgh, 
the Eidgen{\"o}ssische Technische Hochschule (ETH) Z{\"u}rich, 
Fermi National Accelerator Laboratory, the University of Illinois at Urbana-Champaign, the Institut de Ci{\`e}ncies de l'Espai (IEEC/CSIC), 
the Institut de F{\'i}sica d'Altes Energies, Lawrence Berkeley National Laboratory, the Ludwig-Maximilians Universit{\"a}t M{\"u}nchen and the associated Excellence Cluster Universe, 
the University of Michigan, NSF NOIRLab, the University of Nottingham, The Ohio State University, the University of Pennsylvania, the University of Portsmouth, 
SLAC National Accelerator Laboratory, Stanford University, the University of Sussex, Texas A\&M University, and the OzDES Membership Consortium.

Based in part on observations at NSF Cerro Tololo Inter-American Observatory at NSF NOIRLab (NOIRLab Prop. ID 2012B-0001; PI: J. Frieman), which is managed by the Association of Universities for Research in Astronomy (AURA) under a cooperative agreement with the National Science Foundation.

The DES data management system is supported by the National Science Foundation under Grant Numbers AST-1138766 and AST-1536171.
The DES participants from Spanish institutions are partially supported by MICINN under grants PID2021-123012, PID2021-128989 PID2022-141079, SEV-2016-0588, CEX2020-001058-M and CEX2020-001007-S, some of which include ERDF funds from the European Union. IFAE is partially funded by the CERCA program of the Generalitat de Catalunya.

We  acknowledge support from the Brazilian Instituto Nacional de Ci\^encia
e Tecnologia (INCT) do e-Universo (CNPq grant 465376/2014-2).

This document was prepared by the DES Collaboration using the resources of the Fermi National Accelerator Laboratory (Fermilab), a U.S. Department of Energy, Office of Science, Office of High Energy Physics HEP User Facility. Fermilab is managed by Fermi Forward Discovery Group, LLC, acting under Contract No. 89243024CSC000002.

\begin{figure}
    \centering
    \includegraphics[width=\linewidth]{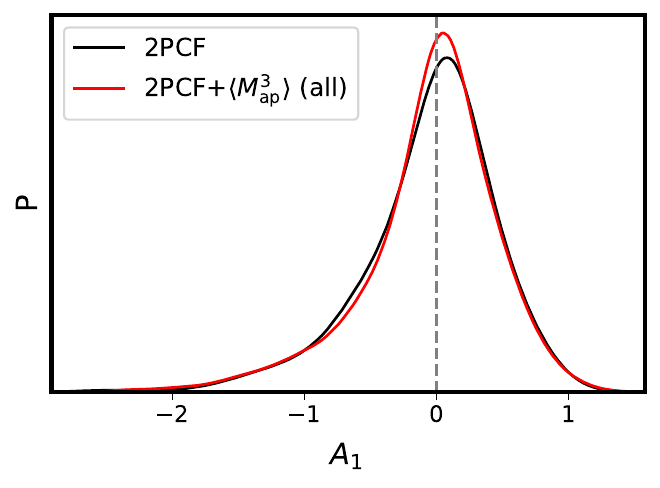}
    \caption{Result of our simulated analysis focusing on the  IA parameter $A_1$. The black and red contours are from 2PCF-only and 2PCF+$\mapcube$ analyses respectively. The vertical dashed line is the input value. A small tightening of the posterior distribution is visible, corresponding to an improvement of 3\% on the 68\% confidence interval of the $A_1$ parameter.}
    \label{fig:chain_test_IA}
\end{figure}

\end{acknowledgments}

\appendix
\section{The BiHalofit formula}\label{sec:bihalofit}

Here we follow the prescription detailed by \citet{Takahashi.Shirasaki.2019} to implement the fitted non-linear bispectrum. To start the BiHalofit computation of the bispectrum, we calculate the following from CAMB \cite{Lewis.Lewis}:

\begin{itemize}
\item The linear matter power spectrum $P(k, z)$ on a grid of k and z values.
\item The non-linear scale $k_{nl}$ as for each redshift value.
\item The comoving distance $\chi(z)$ and the derivative $dz/d\chi (z)$.
\end{itemize}

Next, we compute the fitting parameters, as functions of $\log{\sigma_8}$ and of $n_{\text{eff}}$. We divide the parameters into global and dependent parameters, the former being k-independent, and the latter k-dependent. The global parameters are \cite{Takahashi.Shirasaki.2019}:

\begin{equation*}
\begin{split}
\log{b_n} = -3.428 -2.681\log{\sigma_8} + \\1.624\log{\sigma_8}^2 - 0.095 \log{\sigma_8}^3 \\ \log{c_n} = 0.159 -1.107n_{\text{eff}} \\ \log{\gamma_n} = 0.182 +0.57n_{\text{eff}} \\ \log{f_n} = -10.533-16.838n_{\text{eff}} -\\9.3048 n_{\text{eff}}^2 - 1.8263n_{\text{eff}}^3 
\\ \log{g_n} = 2.787+2.405n_{\text{eff}} +0.4577 n_{\text{eff}}^2 \\ \log{h_n} = -1.118-0.394n_{\text{eff}} \end{split}
\end{equation*}
\begin{equation}
\begin{split}\log{m_n} = -2.605 -2.434\log{\sigma_8} + 5.710\log{\sigma_8}^2 
\\ \log{n_n} = -4.468 -3.080\log{\sigma_8} + 1.035\log{\sigma_8}^2 \\ \log{\mu_n} = 15.312+22.977n_{\text{eff}} +\\10.9579n_{\text{eff}}^2+1.6586n_{\text{eff}}^3  
\\ \log{\nu_n} = 1.347+1.246n_{\text{eff}} +0.4525n_{\text{eff}}^2 \\ \log{p_n} = 0.071-0.433n_{\text{eff}} \\ \log{d_n} = -0.483+0.892\log{\sigma_8} -0.086\Omega_{\rm m}  \\ \log{e_n} = -0.632+0.646n_{\text{eff}} 
\end{split}
\end{equation}

In order to compute the dependent parameters, we sort each set of $(k_1, k_2, k_3)$ to find the set $(k_{\text{min}}, k_{\text{mid}}, k_{\text{max}})$. Then, we define $r_1 \equiv k_{\text{min}}/k_{\text{max}}$ and $r_2=(k_{\text{mid}}+k_{\text{min}}-k{\text{max}})/k_{\text{max}}$. Now we compute the dependent parameters \cite{Takahashi.Shirasaki.2019}:

\begin{equation}
\begin{split}
\log{a_n} = -2.167 - 2.944\log{\sigma_8}-\\1.106\log{\sigma_8}^2 - 2.865\log{\sigma_8}^3-0.310r_1\gamma_n \\ \log{\alpha_n} = min(-4.348 - 3.006n_{\text{eff}} - 0.5745n_{\text{eff}}^2 + \\10^{-0.9+0.2n_{\text{eff}}}r_2^2, n_s) \\
\log{\beta_n} = -1.731 - 2.845n_{\text{eff}} - 1.4995n_{\text{eff}}^2 - \\0.2811n_{\text{eff}}^3 + 0.007r_2
\end{split}
\end{equation}

From these parameters, we compute the one-halo term:

\begin{equation}
B_{1h}(k_1, k_2, k_3,z) = \prod_{i=1,2,3}{\frac{1}{a_n q_i^{\alpha_n} + b_n q_i^{\beta_n}}}\frac{1}{1+(c_n q_i)^{-1}}
\end{equation}

where $q_i = k_i/k_{nl}$.

The 2-halo and 3-halo effects are put together in the biHalofit model $B_{3h}$ term \cite{Takahashi.Shirasaki.2019}. To compute this part, we take the $F_2$ kernel defined as

\begin{equation}
\begin{split}
F_2(k_1,k_2,k_3) = \frac{5}{7} + \frac{2}{7k_1^2k_2^2}\left(\frac{-k_3^2+k_1^2+k_2^2}{2}\right)^2 - \\ \frac{-k_3^2+k_1^2+k_2^2}{2}\left(\frac{1}{2k_1^2}+\frac{1}{2k_2^2}\right)
\end{split}
\end{equation}

Next, we compute the enhanced power spectrum:

\begin{equation}
\begin{split}
P_{\text{enhanced}}(k_i, z) = \frac{1+f_n q_i^2}{1+g_n q_i + h_n q_i^2}P_L(k_i, z) +\\ \frac{1}{m_n q_i^{\mu_n}+n_n q_i^{\nu_n}}\frac{1}{1+(p_nq_i)^{-3}}
\end{split}
\end{equation}

Finally, we get

\begin{equation}
\begin{split}
B_{3h}(k_1, k_2, k_3, z) = 2(F_2(k_1,k_2,k_3) + d_n q_3)\\ \prod_{i=1,2,3}\frac{1}{1+e_n q_i}P_{\text{enhanced}}(k_1,z)P_{\text{enhanced}}(k_2,z)\\ + \text{two permutations}
\end{split}
\end{equation}

and

\begin{equation}
B(k_1, k_2, k_3, z)=B_{1h}(k_1, k_2, k_3, z)+B_{3h}(k_1, k_2, k_3, z)
\end{equation}

\section{$\Lambda$CDM and wCDM mass aperture emulators}
\label{sec:appendix-emu}
Here we specify the configurations of the emulators described in \ref{sec:nn_emulator}. For both $\Lambda$CDM and wCDM, we emulate the redshift-dependent $\mapcube(\theta, z)$ at a specified set of $z$ values. In order to determine the ideal set, we examined possible spacings and attempted to minimize the error relative to our fiducial 130-bin computation. We found that spacing the values linearly between 0.1 and 1.5 with a step size of $\Delta z=0.05$ captured most of the information from this redshift range. We added to the set seven initial values at 0.005, 0.01, 0.02, 0.03, 0.04, 0.05, and 0.07, in order to capture the low-redshift information. This yields us a total of 36 z values.

We then used this set of z values to generate the emulator training set. For $\Lambda$CDM, we used 1300 samples for training/validation and 171 for testing. For wCDM, we used 1700 samples for training/validation and 262 for testing. We perform a logarithmic transformation followed by a min-max scaling as pre-processing steps for both networks.

Finally, we obtain optimized network hyperparameters and train our models. For $\Lambda$CDM, we use 6 hidden layers with 64, 256, 1024, 1024, 384 and 192 nodes, respectively. We perform training with a sequence of decreasing learning rates 
($10^{-2}$, $10^{-3}$, $10^{-4}$, $10^{-5}$, $10^{-6}$), associating them with the progressively reduced batch sizes of 60, 30, 20, 10, and 5. Our final validation loss was equal to $6.36\times 10^{-5}$.

For wCDM, we used 6 hidden layers with 64, 256, 1024, 1024, 256 and 192, respectively. Our learning rate and batch size progressions were equal to that of the $\Lambda$CDM emulator. The final validation loss was equal to $9.83\times 10^{-5}$.

\section{Convergence of the simulated mass aperture covariance}
\label{sec:appendix-convergence}
\begin{figure*}
    \centering
    \includegraphics[width=0.9\linewidth]{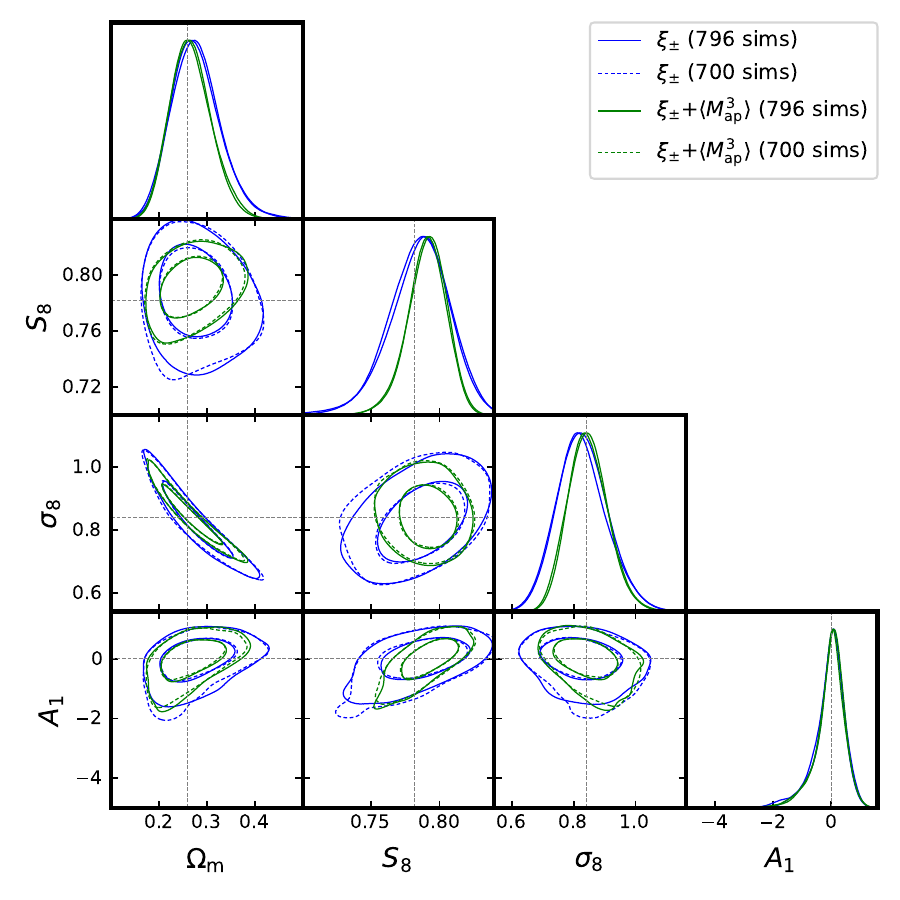}
    \caption{Parameter constraints from $\xi_{\pm}$ and $\mapcube$ using covariance matrices computed with 700 and 796 \software{CosmoGridV1} realizations. Our results are consistent and attest to the convergence of our simulated covariance matrix.}
    \label{fig:robustness_cov}
\end{figure*}

Here we demonstrate the convergence of our simulated covariance matrix. We select 700 out of our 796 \software{CosmoGridV1} realizations to compute an alternative covariance matrix for the full $\xi_{\pm}$ and $\mapcube$ data vector. We compare it with our fiducial matrix and find both to be similar in structure and in magnitude. Next, we perform parameter estimation with the alternate covariance for both the $\xi_{\pm}$-only scenario and the $\xi_{\pm}$+$\mapcube$, following our MOPED compression scheme for the $\xi_{\pm}$ section of the data vector. We find only a minimal alteration of the contours and posteriors in both scenarios. Therefore, we consider that our set of simulations is large enough to accurately describe the covariance of our full data vector. Our results are shown in Figure~\ref{fig:robustness_cov}.

\section{Contours for all cosmological parameters}
\label{sec:allcontours}
\begin{figure*}
    \centering
    \includegraphics[width=\linewidth]{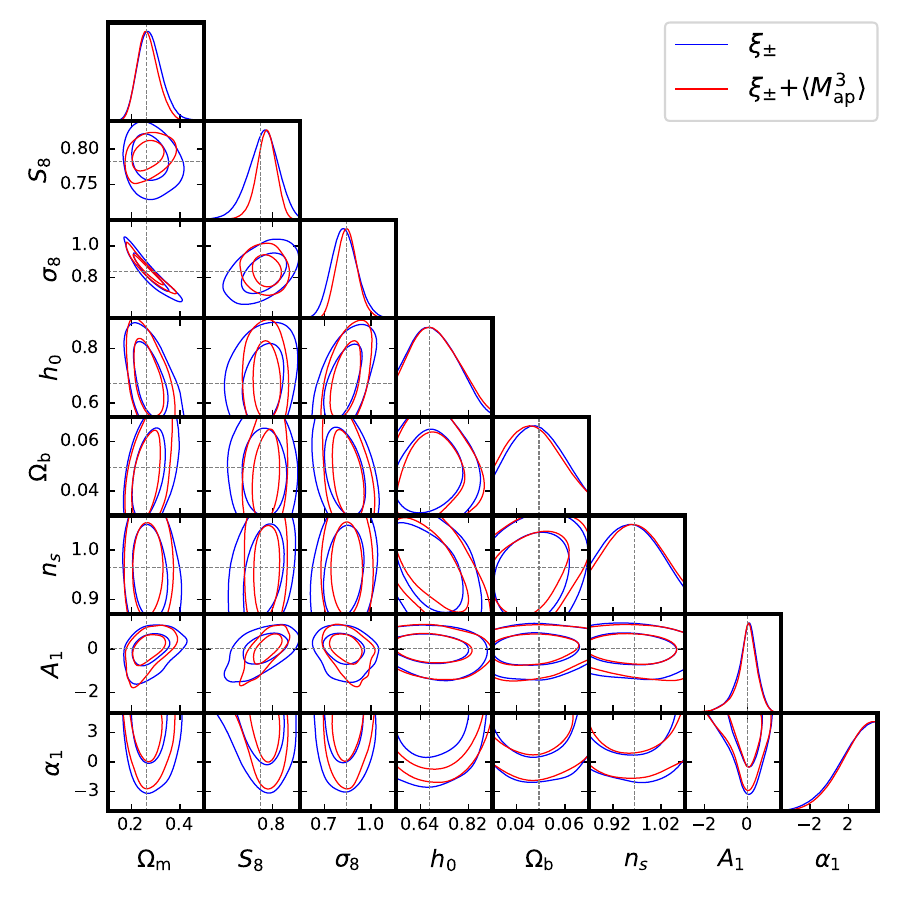}
    \caption{Parameter constraints on the whole set of cosmological parameters from $\xi_{\pm}$ and $\mapcube$ using $\Lambda$CDM modeling. The blue contours indicate results from the 2PCF alone, while the red contours indicate results from the 2PCF combined with the mass aperture statistic. The gray dotted lines represent the input values for the parameters. We do not indicate any value for $\alpha_1$ since our input data vector had a zero intrinsic alignment amplitude.}
    \label{fig:lcdm_allparams}
\end{figure*}

\begin{figure*}
    \centering
    \includegraphics[width=\linewidth]{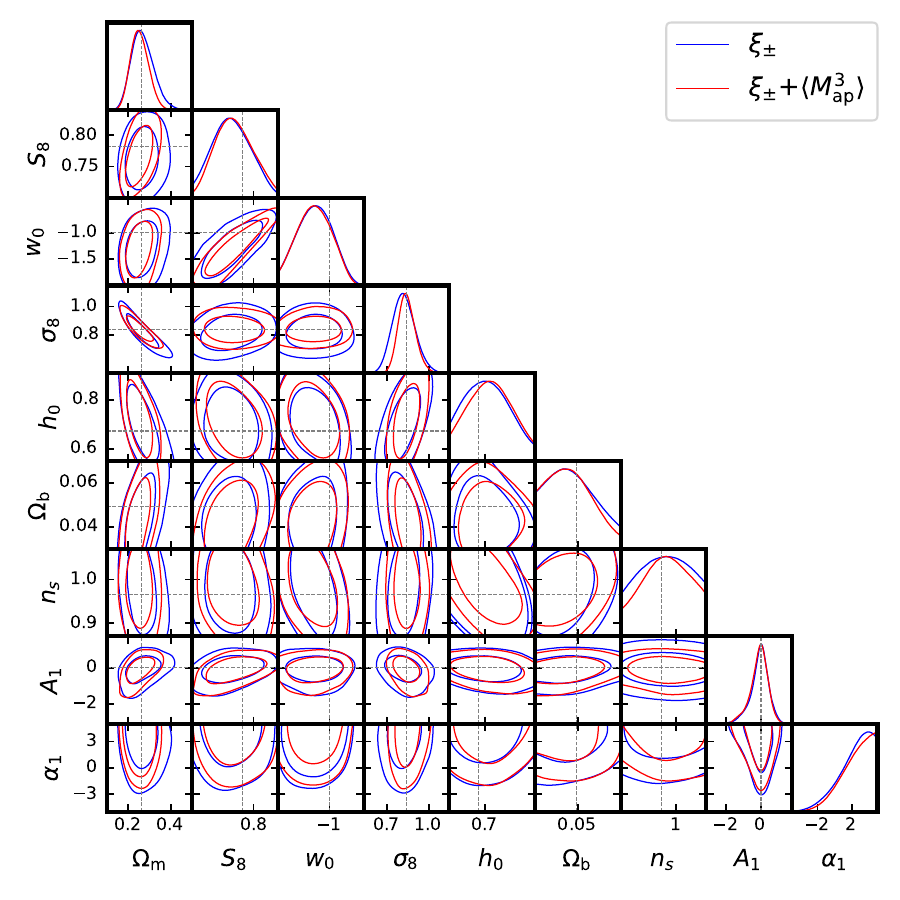}
    \caption{Parameter constraints on the whole set of cosmological parameters from $\xi_{\pm}$ and $\mapcube$ using wCDM modeling. The blue contours indicate results from the 2PCF alone, while the red contours indicate results from the 2PCF combined with the mass aperture statistic. The gray dotted lines represent the input values for the parameters. As in the previous figure, we do not indicate any value for $\alpha_1$ since our input data vector had a zero intrinsic alignment amplitude.}
    \label{fig:wcdm_allparams}
\end{figure*}

Here we present our $\Lambda$CDM and wCDM contours for the full set of cosmological parameters. We exclude the constraints on the sum of the neutrino masses because our $\mapcube$ emulator assumes a fixed value of $\Sigma m_{\nu}=0.06$. For $\Lambda$CDM, our results are shown in Figure~\ref{fig:lcdm_allparams}. For wCDM, we present our results in Figure~\ref{fig:wcdm_allparams}.

\bibliography{refs}

\end{document}